\documentclass[12pt,letterpaper]{article}
\usepackage{epsfig,rotating,setspace,latexsym,amsmath,epsf,amssymb,amsfonts,bm,theorem,cite,caption,subcaption,enumerate,longtable,accents}
\usepackage{algorithm,algorithmic,graphicx,epsf,authblk,epstopdf,url,color,multirow}
\usepackage{mathtools}
\usepackage{comment}
\usepackage{graphicx}
\usepackage{amsmath}

\newtheorem{theorem}{Theorem}

\newtheorem{remark}{Remark}

\allowdisplaybreaks

\begin{document}

\title{Private Read-Update-Write with Controllable Information Leakage for Storage-Efficient Federated Learning with Top $r$ Sparsification}

\author{Sajani Vithana \qquad Sennur Ulukus\\
	\normalsize Department of Electrical and Computer Engineering\\
	\normalsize University of Maryland, College Park, MD 20742 \\
	\normalsize {\it spallego@umd.edu} \qquad {\it ulukus@umd.edu}}

\date{}
\maketitle

\vspace*{-1cm}

\begin{abstract}
In federated learning (FL), a machine learning (ML) model is collectively trained by a large number of users, using their private data in their local devices. With top $r$ sparsification in FL, the users only upload the most significant $r$ fraction of updates, and the servers only send the most significant $r'$ fraction of parameters to the users in order to reduce the communication cost. However, the values and the indices of the sparse updates leak information about the users' private data. In this work, we consider an FL setting where $N$ non-colluding databases store the model to be trained, from which the users download and update sparse parameters privately, without revealing the values of the updates or their indices to the databases. We propose four schemes with different properties to perform this task while achieving the minimum communication costs, and show that the information theoretic privacy of both values and positions of the sparse updates can be guaranteed. This is achieved at a considerable storage cost, though. To alleviate this, we generalize the schemes in such a way that the storage cost is reduced at the expense of a certain amount of information leakage, using a model segmentation mechanism. In general, we provide the tradeoff between 
communication cost, storage cost and information leakage in private FL with top $r$ sparsification. 
\end{abstract}

\section{Introduction}

Federated learning (FL) \cite{FL1,FL2} is a widely used distributed learning technique where a set of users remotely train a ML model using their own local data in their own devices, and share only the gradient updates with the central server. This reduces the privacy leakage of data providers while decentralizing the processing power requirements of the central server. However, it has been shown that the gradients shared by a user can be used to obtain information about the user's private data \cite{comprehensive, MembershipInterference, featureLeakage, SecretSharer, InvertingGradients, DeepLeakage, BeyondClassRepresentatives}. Cryptographic protocols as in secure aggregation \cite{PracticalSecureAgg}, differential privacy (DP) \cite{DP} via noise addition, data sampling and data shuffling, e.g., \cite{reinforcement, avgDP, cpSGD, PrivacyAmp, cross, language, DPFL, shuffle, PrivacyBlanket, shuffledDPFL, client, recent, aggregation} are some of the methods used to minimize this information leakage in FL. However, these methods do not guarantee information theoretic privacy of each individual user's local data. 

Apart from the privacy leakage, another drawback of FL is the large communication cost incurred by sharing model parameters and updates with millions of users in multiple rounds. Some of the solutions to this problem include, gradient quantization \cite{qsl,fedpaq,qsgd,constraints}, federated submodel learning (FSL) \cite{billion, paper1, secureFSL, dropout, ourICC, rw_jafar, pruw, sparse, rd}, and gradient sparsification \cite{sparse1, GGS, adaptive, conv, conv2, overtheair, rtopk, timecorr}. In gradient quantization, the values of the gradients are quantized and represented with a fewer number of bits. In FSL, the ML model is divided into multiple submodels based on different types of data used to train the entire model, and each user only downloads and updates the submodel that can be updated by its own local data. In gradient sparsification, the users only communicate a selected set of gradients and parameters as opposed to communicating all gradient updates and parameters. Typically the sparsification rates (fraction of the parameters/updates communicated) are around $10^{-2}$ to $10^{-3}$, which significantly reduces the communication cost. 

Top $r$ sparsification is a widely used sparsification technique, where only the most significant $r$ fraction of parameters/updates are shared between the users and the central server. In certain cases, it has been shown that top $r$ sparsification outperforms classical FL. However, the values as well as the positions (indices) of the sparse updates leak information about the user's local data. Note that the positions of the sparse updates leak information about the most and least significant sets of parameters for a given user, which can be used to infer information about the user's private data. Thus, in order to guarantee the privacy of users participating in the sparse FL process, two components need to be kept private, namely, 1) values of sparse updates, 2) indices of sparse updates. In this work, we develop schemes to perform the user-database communications in FL with top $r$ sparsification while guaranteeing information theoretic privacy of the values and indices of the sparse updates. 

\begin{figure*}[t]
    \centering
    \includegraphics[scale=0.75]{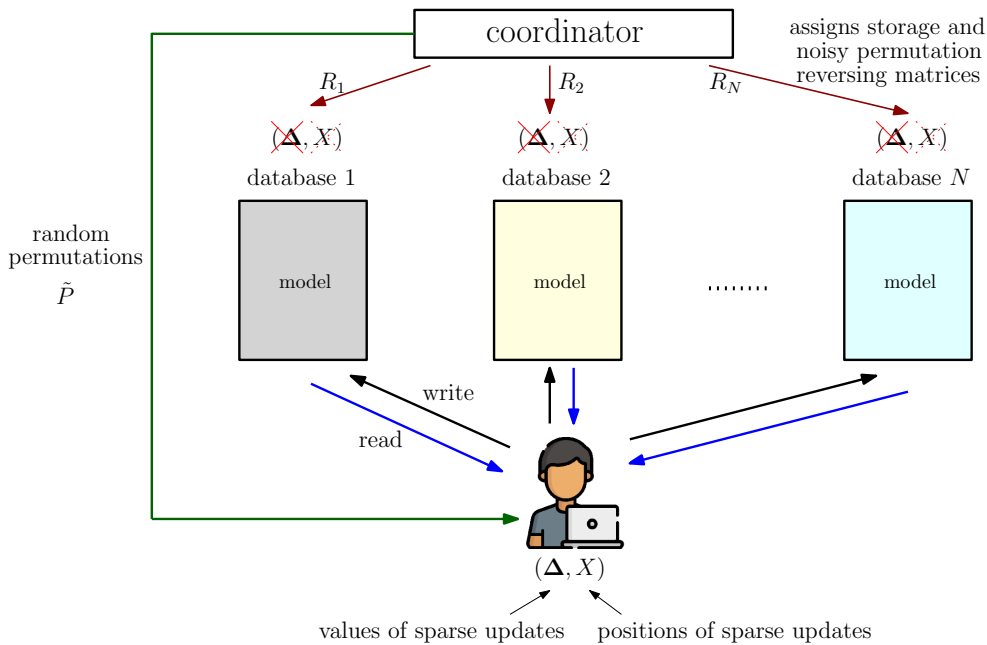}
    \caption{System model: A user reads (downloads), updates, writes (uploads) a ML model.}
    \label{model}
\end{figure*}

\begin{figure}[t]
    \centering
    \begin{subfigure}[b]{0.75\textwidth}
        \centering
        \includegraphics[width=\textwidth]{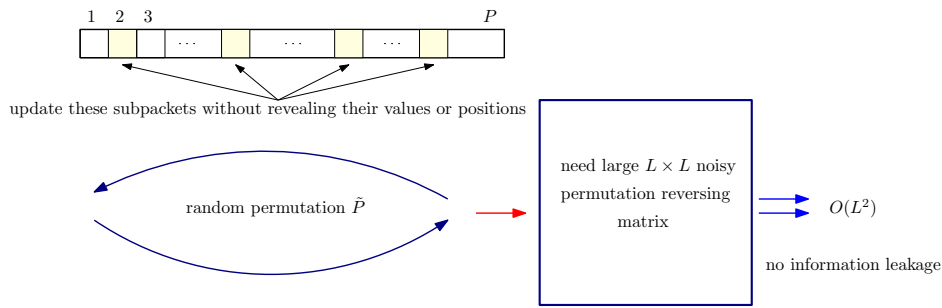}
        \caption{Without segmentation.}
        \label{fig:read}
    \end{subfigure}
    \vfill \vspace*{0.5cm}
    \begin{subfigure}[b]{0.75\textwidth}
        \centering
        \includegraphics[width=\textwidth]{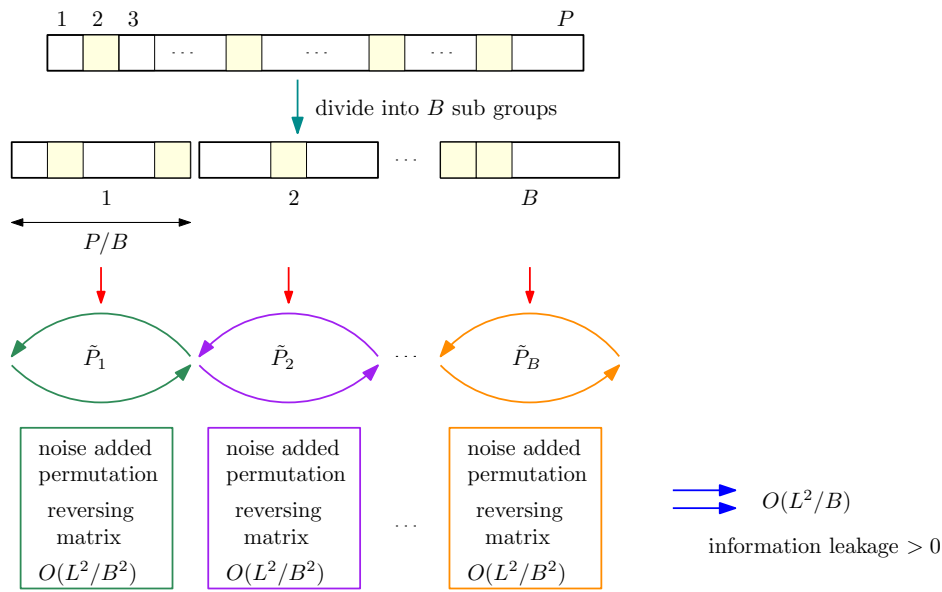}
        \caption{With segmentation.}
        \label{fig:write}
    \end{subfigure}
    \caption{Motivation for segmentation in permutation techniques: (a) Permutation of the entire model without segmentation. (b) Permutation within segments with segmentation.}
    \label{idea}
\end{figure}

We consider a FL setting with multiple non-colluding databases storing the ML model, and a single user communicating with each of the databases as shown in Fig.~\ref{model}. The schemes we propose in this work are based on permutation techniques, where a coordinator initializes a random permutation of sets of parameters, and sends it to the users. The coordinator then places noise added permutation reversing matrices at each database in such a way that the databases learn nothing about the underlying permutation. All communications between the user and the databases take place in terms of the permuted indices, which guarantee the privacy of the indices of the sparse updates.\footnote{Rigorous proofs on privacy/information leakage are provided in Section~\ref{info_leak}.} However, the parameters in each database get updated in the correct order, with the aid of the noise added permutation reversing matrices. The main drawback of this method is the considerably large storage cost incurred by the large permutation reversing matrices. To that end, we propose schemes that reduce the storage cost by decreasing the size of the noise added permutation reversing matrices, at the expense of a given amount of information leakage. This is achieved by dividing the ML model into multiple segments and carrying out permutations within each segment. This is illustrated in Fig.~\ref{idea}. The number of segments is chosen based on the allowed amount of information leakage and the storage capacity of the databases.

In this work, we propose four schemes to perform user-database communications in private FL with top $r$ sparsification with different properties such as lower communication costs, lower storage costs or lower amounts of information leakage. The four schemes differ from each other based on the storage structure (MDS coded or uncoded) and the permutation mechanism (only within-segment permutations or within and inter-segment permutations) used. MDS coded storage decreases the storage cost while increasing the communication cost, and the two-stage permutations (within and inter-segment permutations) decrease the information leakage significantly compared to single-stage permutations (only within-segment permutations), while slightly increasing the communication cost. Based on the specifications and limitations of the given FL task, one can choose the most suitable scheme. In general, we present the tradeoff between the communication cost, storage complexity and information leakage in private FL with sparsification.

\section{Problem Formulation}\label{problem}

We consider a FL setting in which a ML model consisting of $L$ parameters belonging to $P$ subpackets is stored in $N$ non-colluding databases. The parameters take values from a large enough finite field $\mathbb{F}_q$. A given user at a given time $t$ reads (downloads) the required parameters of the model from the databases, trains the model using the user's local data, and writes (uploads) the most significant $r$ fraction of updates back to all databases. In this work, we consider sparsification in both uplink and downlink, to reduce the communication cost. In particular, the sparsification rates of the reading (downlink) and writing (uplink) phases are given by $r'$ and $r$, respectively. In other words, in the reading (download) phase, the users only download a selected set of $Pr'$ subpackets determined by the databases.\footnote{These subpackets could be determined by the databases based on the sparse updates received at the previous time step, or by any other downlink sparsification protocol. For example, the databases can choose the most commonly updated $Pr'$ subpackets in the writing phase of time $t-1$ to be sent to the users in the reading phase at time $t$.} Once the model is trained locally, each user only uploads the most significant $Pr$ set of updates (corresponding subpackets) to the databases in the writing (upload) phase.\footnote{We assume that all values in the sparse set of $Pr$ subpackets in the writing phase are non-zero.}

Note that  the users send no information to the databases in the reading phase. Therefore, no information about the user's local data is leaked to the databases in the reading phase. The users send the sparse updates and their positions (indices) to the databases in the writing phase to train the model. Information about the user's local data can be leaked to the databases from these updates and their indices.\footnote{The positions (indices) of the sparse updates leak information about the most and least significant parameters in the model for a given user, which may leak information about the user's local data.}
In this work, we consider the following privacy guarantees on the values and the indices of the sparse updates.

\emph{Privacy of the values of sparse updates:} No information on the values of the sparse updates is allowed to leak to any of the databases, i.e., 
\begin{align}
    I(\Delta_i^{[t]};G_n^{[t]})=0, \quad n\in\{1,\dotsc,N\}, \quad \forall i
\end{align}
where $\Delta_{i}^{[t]}$ is the value of the $i$th sparse (non-zero) update of a given user at time $t$ and $G_{n}^{[t]}$ contains all the information sent by the user to database $n$ at time $t$.

\emph{Privacy of the positions (indices) of sparse updates:} The amount of information leaked on the indices of the sparse updates need to be maintained under a given privacy leakage budget $\epsilon$, i.e.,
\begin{align}
    I(X^{[t]};G_n^{[t]})\leq\epsilon, \quad n\in\{1,\dotsc,N\}, 
\end{align}
where $X^{[t]}$ is the set of indices of the sparse subpackets updated by a given user at time $t$. The system model with the privacy constraints is shown in Fig.~\ref{model}. A coordinator is used to initialize the FL process.\footnote{The coordinator is only available at the initialization stage, and will not be part of the system model once the FL process begins.} In addition to the privacy constraints, we require the following security and correctness conditions for the reliability of the scheme.

\emph{Security of the model:} No information about the model parameters is allowed to leak to the databases, i.e.,
\begin{align}
    I(W^{[t]};S_n^{[t]})=0, \quad n\in\{1,\dotsc,N\},
\end{align}
where $W^{[t]}$ is the ML model and $S_n^{[t]}$ is the data content in database $n$ at time $t$.

\emph{Correctness in the reading phase:} The user should be able to correctly decode the sparse set of subpackets (denoted by $J$) of the model, determined by the downlink sparsification protocol, from the downloads in the reading phase, i.e., 
\begin{align}
H(W_{J}^{[t-1]}|A_{1:N}^{[t]})=0,
\end{align}
where $W_{J}^{[t-1]}$ is the set of subpackets in  set $J$ of the model $W$ at time $t-1$ (before updating) and $A_n^{[t]}$ is the information downloaded from database $n$ at time $t$.

\emph{Correctness in the writing phase:} Let $J'$ be the set of most significant $Pr$ subpackets of the model, updated by a given user at time $t$. The model should be correctly updated as,
\begin{align}
    W_{s}^{[t]}=
    \begin{cases}
    W_{s}^{[t-1]}+\Delta_{s}^{[t]}, & \text{if $s\in J'$}\\
    W_{s}^{[t-1]}, & \text{if $s\notin J'$}
    \end{cases},
\end{align}
where $W_{s}^{[t-1]}$ is subpacket $s$ of the model at time $t-1$ and $\Delta_{s}^{[t]}$ is the corresponding update of subpacket $s$ at time $t$.

\emph{Reading and writing costs:} The reading and writing costs are defined as $C_R=\frac{\mathcal{D}}{L}$ and $C_W=\frac{\mathcal{U}}{L}$, respectively, where $\mathcal{D}$ is the total number of symbols downloaded in the reading phase, $\mathcal{U}$ is the total number of symbols uploaded in the writing phase, and $L$ is the size of the model. The total cost $C_T$ is the sum of the reading and writing costs $C_T=C_R+C_W$.

\emph{Storage complexity:} The storage complexity is quantified by the order of the total number of symbols stored in each database. 

In this work, we propose schemes to perform FL with top $r$ sparsification, that result in the minimum total communication costs and storage complexities, while satisfying all privacy, security and correctness conditions described above.

\section{Main Result}\label{main}

\begin{theorem}\label{main_result}
Consider a FL model stored in $N$ non-colluding databases, consisting of $L$ parameters with values from a finite field $\mathbb{F}_q$, which are included in $P$ subpackets. The model is divided into $B$ segments of equal size ($1\leq B<P$), such that each consecutive $\frac{P}{B}$ subpackets constitute each segment. Assume that the FL model is being updated by users at each time instance with uplink and downlink sparsification rates (top $r$ sparsification) of $r$ and $r'$, respectively. Let $\hat{X}_i$ be the random variable representing the number of subpackets with sparse (non-zero) updates in the $i$th segment, uploaded by any given user, and let $(\Tilde{X}_1,\dotsc,\Tilde{X}_B)$ be the general vector representing all distinct combinations of $(\hat{X}_1,\dotsc,\hat{X}_B)$, irrespective of the segment index. Then, the reading/writing costs, storage complexities and amounts of information leakage presented in Table~\ref{main_res} are achievable in a single round of the FL process in the perspective of a single user.

\begin{table}[h]
\begin{center}
\begin{tabular}{ |c|c|c|c|c| }
\hline
  case & reading cost & writing cost & storage complexity & information leakage\\ 
  \hline
  1 & $\frac{2r'(1+\frac{\log_q P}{N})}{1-\frac{2}{N}}$ & $\frac{2r(1+\log_q P)}{1-\frac{2}{N}}$ & $O(\frac{L^2}{B})$ & $H(\hat{X}_1,\dotsc,\hat{X}_B)$\\ 
  \hline
  2 & $\frac{3r'(1+\frac{\log_q P}{N})}{1-\frac{1}{N}}$ & $\frac{3r(1+\log_q P)}{1-\frac{1}{N}}$ & $O(\frac{L^2}{BN^2})$ & $H(\hat{X}_1,\dotsc,\hat{X}_B)$\\
  \hline
  3 & $\frac{2r'(1+\frac{\log_q P}{N})}{1-\frac{4}{N}}$ & $\frac{2r(1+\log_q P)}{1-\frac{4}{N}}$ & $\max\{O(\frac{L^2}{B}),O(N^2B^2)\}$ & $H(\Tilde{X}_1,\dotsc,\Tilde{X}_B)$\\
  \hline
  4 & $\frac{5r'(1+\frac{\log_q P}{N})}{1-\frac{1}{N}}$ & $\frac{5r(1+\log_q P)}{1-\frac{1}{N}}$ & $\max\{O(\frac{L^2}{N^2B}),O(B^2)\}$ & $H(\Tilde{X}_1,\dotsc,\Tilde{X}_B)$\\
  \hline
\end{tabular}
\end{center}
\vspace*{-0.2cm}
\caption{Achievable sets of communication costs, storage costs and information leakage.}
\label{main_res}
\end{table}
\end{theorem}

\begin{remark}
    The information leakage in Table~\ref{main_res} corresponds to the amount of information leaked on the indices of the sparse updates.\footnote{Information theoretic privacy of the values of updates is guaranteed, as stated in the problem formulation.} For a given privacy leakage budget on the indices of the sparse updates given by $\epsilon$, the optimum number of segments $B$ can be calculated by minimizing the storage complexity, such that $H(\hat{X}_1,\dotsc,\hat{X}_B)<\epsilon$ or $H(\tilde{X}_1,\dotsc,\Tilde{X}_B)<\epsilon$ is satisfied (based on the considered case). This is valid for all four cases.
\end{remark}

\begin{remark}
When $B=1$ (no segmentation present), $\hat{X}_1=\Tilde{X}_1=Pr$ and the corresponding information leakage is zero since $Pr$ is fixed and $H(\hat{X}_1)=H(\Tilde{X}_1)=0$, i.e., the four schemes corresponding to the four cases achieve information theoretic privacy of the values and positions of the sparse updates while incurring the same communication costs stated in Table~\ref{main_res}, when $B=1$. However, in this case, the storage costs increase to either $O(L^2)$ or $O\left(\frac{L^2}{N^2}\right)$.
\end{remark}

\begin{remark}    $H(\hat{X}_1,\dotsc,\hat{X}_B)>H(\Tilde{X}_1,\dotsc,\Tilde{X}_B)$ since $H(\hat{X}_1,\dotsc,\hat{X}_B)$ considers all possible values of $\hat{X}_i$, while $H(\Tilde{X}_1,\dotsc,\Tilde{X}_B)$ only considers distinct sets of $\{\hat{X}_i\}_{i=1}^B$. For example, if $B=2$, $H(\hat{X}_1,\hat{X}_2)$ considers both permutations $\{1,2\}$ and $\{2,1\}$ while $H(\Tilde{X}_1,\Tilde{X}_2)$ only takes one of them into account, i.e., the probabilities considered in $H(\Tilde{X}_1,\dotsc,\Tilde{X}_B)$ are more dense and concentrated compared to that of $H(\hat{X}_1,\dotsc,\hat{X}_B)$.
\end{remark}

\begin{remark}
    Cases 1-4 are achieved by schemes that utilize both CSA \cite{CSA} and permutation techniques which are described in detail in Section~\ref{proposedscheme}. The schemes for cases 1 and 2 use a single round permutation technique (only within-segment permutations) while cases 3 and 4 use a two-round permutation technique (both within and inter-segment permutations) which reduces the information leakage further. Cases 3 and 4 are extensions of cases 1 and 2, respectively, with the additional permutation round. Cases 3 and 4 incur larger communication costs compared to cases 1 and 2, while resulting in lower amounts of information leakage.
\end{remark}

\begin{remark}
    The four cases (schemes) have different properties. Cases 1 and 3 result in the lowest communication costs at the expense of a larger storage complexity resulted by replicated storage and larger noisy permutation reversing matrices. Cases 2 and 4 use MDS coded storage and compact permutation reversing matrices, which reduces the storage complexity at the expense of larger communication costs. 
\end{remark}

\begin{remark}
    The communication cost does not depend on the number of segments $B$.
\end{remark}

\begin{figure}[t]
    \centering
    \includegraphics[scale=0.7]{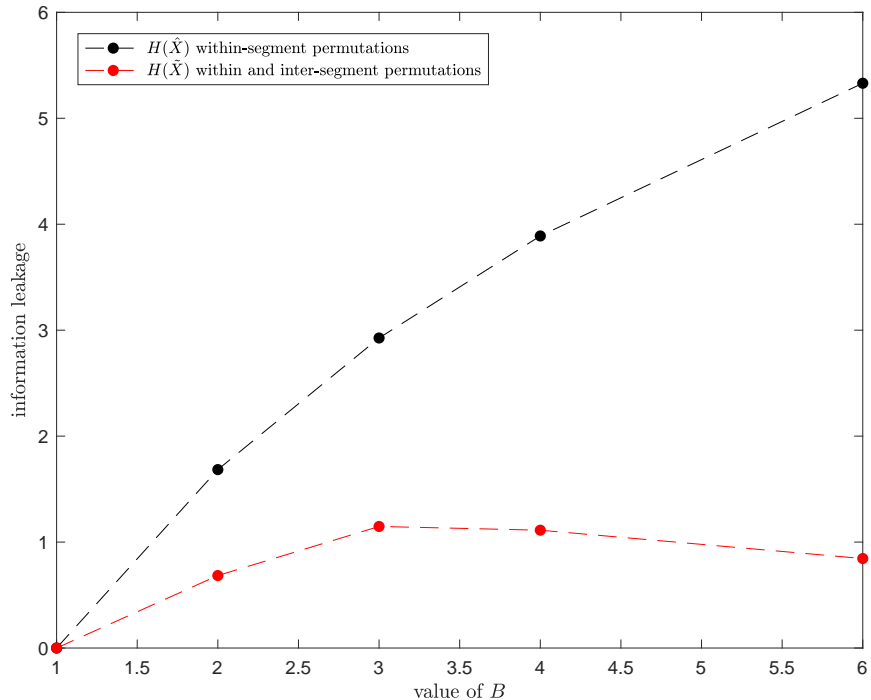}
    \caption{Information leakage of an example setting with $P=12$ and different values of $B$.}
    \label{leak}
\end{figure}

\begin{remark}
    Consider an example setting with $P=12$ subpackets divided into $B=1,2,3,4,6$ segments. Assume that each subpacket is equally probable to be selected to the set of most significant $Pr=3$ subpackets. The behavior of the information leakage for each value of $B$ is shown in Fig.~\ref{leak}.
\end{remark}

\section{Proposed Schemes}\label{proposedscheme}

\subsection{General Schemes With Examples}

\begin{figure}
		\centering
		\includegraphics[scale=1]{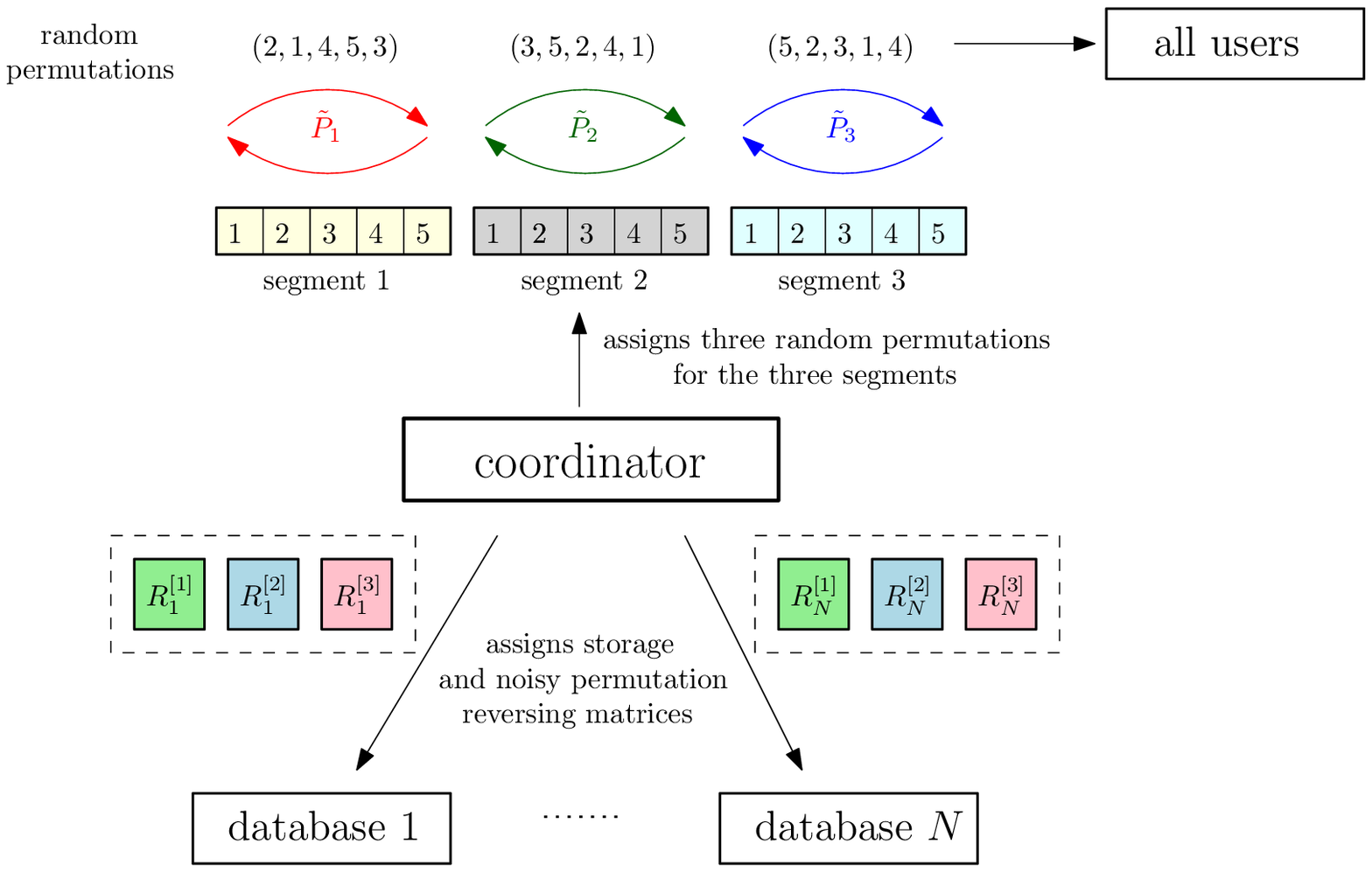}
		\caption{Initialization of the scheme for cases 1 and 2.}
		\label{init_c1_eg}
\end{figure}

In this section, we provide the proposed schemes for all four cases. In all four schemes, we divide the $P$ subpackets into $B$ non-overlapping equal-sized segments to control the storage cost and the information leakage. The parameter $B$ is a variable that can be chosen based on the given privacy leakage budget and the limitations on the storage capacities. In this section, we present the general schemes for arbitrary values of $B$, $P$, $r$ and $r'$. As a further illustration, we provide examples along with the general scheme for all four cases. In the two examples corresponding to cases 1 and 2, we assume the same setting with $P=15$ subpackets (subpacketization $\ell$), divided into $B=3$ equal segments as shown in Fig.~\ref{init_c1_eg}.

\textbf{Case 1:} Uncoded\footnote{Even though the model parameters and noise symbols are combined together (coded form) in the storage in \eqref{storage_c1_eg}, each parameter is not combined with other parameters, resulting in uncoded storage.} storage and larger permutation reversing matrices are used in this case to reduce the communication cost, at the expense of a larger storage cost.

\emph{Initialization:} A single subpacket (subpacket $s$) in case 1 is stored in database $n$, $n\in\{1,\dotsc,N\}$ as,
\begin{align}\label{storage_c1_eg}
 S_n^{[s]}=\begin{bmatrix}
    \frac{1}{f_{1}-\alpha_n}W_{1}^{[s]}+\sum_{j=0}^{\ell}\alpha_n^j Z_{1,j}^{[s]}\\
    \vdots\\
    \frac{1}{f_{\ell}-\alpha_n}W_{\ell}^{[s]}+\sum_{j=0}^{\ell}\alpha_n^j Z_{\ell,j}^{[s]}
    \end{bmatrix},
\end{align}
where $W_i^{[s]}$ is the $i$th parameter of subpacket $s$, $Z_{i,j}^{[s]}$ are random noise symbols and $\{f_i\}_{i=1}^{\ell}, \{\alpha_n\}_{n=1}^N$ are globally known distinct constants from $\mathbb{F}_q$. The subpackets in each segment are stacked one after the other in the order of subpacket $1$ through subpacket $\frac{P}{B}$. At the initialization stage, the coordinator sends $B$ ($B=3$ for the example considered) randomly and independently chosen permutations of the $\frac{P}{B}$ ($\frac{P}{B}=5$ for the example considered) subpackets in each of the $B$ segments to all users. These permutations are denoted by $\tilde{P}_1,\dotsc,\tilde{P}_B$. The coordinator also sends the $B$ corresponding noise added permutation reversing matrices given by,
\begin{align}
    R^{[i]}_n=(\tilde{R}^{[i]}\otimes\Gamma_n)+\tilde{Z}^{[i]}, \quad i=1,\dotsc,B,
\end{align}
to database $n$, $n\in\{1,\dotsc,N\}$, as shown in Fig.~\ref{init_c1_eg}, where $\tilde{R}^{[i]}$ is the permutation reversing matrix corresponding to the permutation $\tilde{P}_i$, $\Gamma_n$ is the diagonal matrix given by,
\begin{align}
   \Gamma_n=\begin{bmatrix}
    \frac{1}{f_1-\alpha_n} & & \\ & \ddots & \\& & \frac{1}{f_\ell-\alpha_n} 
\end{bmatrix}, 
\end{align}
and $\tilde{Z}^{[i]}$ is a random noise matrix of size $\frac{P\ell}{B}\times\frac{P\ell}{B}$. Based on the example considered, the permutation reversing matrix for database $n$, $n\in\{1,\dotsc,N\}$ corresponding to the first segment (permutation: $\Tilde{P}_1=(2,1,4,5,3)$) is given by,
\begin{align}\label{R1}
    R_n^{[1]}=\left(\begin{bmatrix}
        0 & 1 & 0 & 0 & 0\\
        1 & 0 & 0 & 0 & 0\\
        0 & 0 & 0 & 0 & 1\\
        0 & 0 & 1 & 0 & 0\\
        0 & 0 & 0 & 1 & 0\\
    \end{bmatrix} \otimes \Gamma_n\right)+\tilde{Z}^{[1]}
    =\begin{bmatrix}
        0_{\ell\times\ell} & \Gamma_n & 0_{\ell\times\ell} & 0_{\ell\times\ell} & 0_{\ell\times\ell}\\
        \Gamma_n & 0_{\ell\times\ell} & 0_{\ell\times\ell} & 0_{\ell\times\ell} & 0_{\ell\times\ell}\\
        0_{\ell\times\ell} & 0_{\ell\times\ell} & 0_{\ell\times\ell}  & 0_{\ell\times\ell}& \Gamma_n\\
        0_{\ell\times\ell} & 0_{\ell\times\ell} & \Gamma_n & 0_{\ell\times\ell}&  0_{\ell\times\ell} \\
        0_{\ell\times\ell}  & 0_{\ell\times\ell} & 0_{\ell\times\ell} & \Gamma_n & 0_{\ell\times\ell}\\
    \end{bmatrix}+\tilde{Z}^{[1]},
\end{align}
Similarly, for the second segment (permutation:$\Tilde{P}_2=(3,5,2,4,1)$), the permutation reversing matrix for database $n$, $n\in\{1,\dotsc,N\}$ is given by,
\begin{align}\label{R1}
    R_n^{[2]}=\begin{bmatrix}
        0_{\ell\times\ell} & 0_{\ell\times\ell} & 0_{\ell\times\ell} & 0_{\ell\times\ell}& \Gamma_n\\
        0_{\ell\times\ell} & 0_{\ell\times\ell} & \Gamma_n & 0_{\ell\times\ell} & 0_{\ell\times\ell}\\
        \Gamma_n & 0_{\ell\times\ell} & 0_{\ell\times\ell} & 0_{\ell\times\ell}  & 0_{\ell\times\ell}\\
        0_{\ell\times\ell} & 0_{\ell\times\ell} &  0_{\ell\times\ell}& \Gamma_n & 0_{\ell\times\ell} \\
        0_{\ell\times\ell}  & \Gamma_n & 0_{\ell\times\ell} & 0_{\ell\times\ell} & 0_{\ell\times\ell}\\
    \end{bmatrix}+\tilde{Z}^{[2]}.
\end{align}

The coordinator leaves the system once the storage, permutations and noise added permutation reversing matrices are initialized and the system is ready to begin the FL process. All subsequent communications take place only between individual users and databases in terms of permuted subpacket indices. The databases never learn the underlying permutations despite having access to the noise added permutation reversing matrices, since the added noise $\tilde{Z}^{[i]}$ makes the noisy matrices independent of the original permutation reversing matrix from Shannon's one time pad theorem.

\emph{Reading Phase:} The databases decide the permuted indices of the $Pr'$ sparse subpackets to be sent to the users at time $t$ in the reading phase, based on the permuted subpacket indices received in the writing phase at time $t-1$. For example, the databases consider the permuted indices of the subpackets updated by all users at time $t-1$, and select the most popular $Pr'$ of them to be sent to the users in the reading phase of time $t$. Note that the databases are unaware of the real indices of the sparse subpacket indices updated by users in the writing phase at each time instance and only work with the permuted indices in both phases. We denote the permuted indices of the sparse subpackets to be sent to the users from segment $j$ as $\tilde{V}_j$ for $j\in\{1,\dotsc,B\}$. For this example, let the sparse set of permuted subpacket indices corresponding to the first segment be $\Tilde{V}_1=\{1,3\}$.\footnote{Two similar sets (with same or different cardinalities, such that the sum of all three cardinalities equals $Pr'$) exist for segments 2 and 3 as well.} One designated database sends these permuted indices of each segment to the users. The users then find the real indices, using the known permutations as $V_j(i)=\tilde{P}_j(\tilde{V}_j(i))$ for each sparse subpacket $i$ in segment $j\in\{1,\dotsc,B\}$. For this example (segment 1), the real set of indices is given by,
\begin{align}
    V_1(i)&=\Tilde{P}_1(\Tilde{V}_1(i)), \quad i=1,2\\
    V_1&=\{2,4\}\label{real_id}.
\end{align}
In order to send the $i$th sparse subpacket of segment $j$, $\Tilde{V}_j(i)$, each database $n$, $n\in\{1,\dotsc,N\}$ generates the following query.
\begin{align}
    Q_n^{[\Tilde{V}_j(i)]}=\sum_{k=1}^\ell R_n^{[j]}(:, (i-1)\ell+k).
\end{align}
For example, the query corresponding to the first sparse subpacket of the first segment (i.e., $\Tilde{V}_1(1)=1$) is given by,
\begin{align}
    Q_n^{[\Tilde{V}_1(1)]}=Q_n^{[1]}&=\sum_{k=1}^\ell R_n^{[1]}(:, k)=\begin{bmatrix}
        0_{\ell}\\\frac{1}{f_1-\alpha_n}\\\vdots\\\frac{1}{f_\ell-\alpha_n}\\0_{\ell}\\0_{\ell}\\0_{\ell}\\
    \end{bmatrix}+Z_1,
\end{align}
where $Z_1$ is a random noise vector resulted by the noise component of $R_n^{[1]}$. Similarly, the query for the second sparse subpacket of segment 1 (i.e., $\Tilde{V}_1(2)=3$) is given by,
\begin{align}
    Q_n^{[\Tilde{V}_1(2)]}=Q_n^{[3]}&=\sum_{k=1}^\ell R_n^{[1]}(:, 2\ell+k)=\begin{bmatrix}        0_{\ell}\\0_{\ell}\\0_{\ell}\\\frac{1}{f_1-\alpha_n}\\\vdots\\\frac{1}{f_\ell-\alpha_n}\\0_{\ell}\\
    \end{bmatrix}+Z_2.
\end{align}
Note that the reversal of the permutations is hidden from the databases by the random noise vectors $Z_1$ and $Z_2$. Then, database $n$, $n\in\{1,\dotsc,N\}$ sends the answer corresponding to the $i$th sparse subpacket of segment $j$ to all users by calculating the dot product between the queries and the scaled storage as,
\begin{align}
     A_n^{[\Tilde{V}_j(i)]}&=(D_n\times S_n)^T Q_n^{[\Tilde{V}_j(i)]}, \quad j\in\{1,\dotsc,B\}\\
     &=\frac{1}{f_1-\alpha_n}W_1^{[V_j(i)]}+\dotsc+\frac{1}{f_\ell-\alpha_n}W_\ell^{[V_j(i)]}+P_{\alpha_n}(\ell+1)\label{general_r},
\end{align}
where $D_n$ is the diagonal matrix of size $\frac{P\ell}{B}\times\frac{P\ell}{B}$ given by,
\begin{align}
   D_n=I_{\frac{P}{B}}\otimes \Gamma_n^{-1}=
    \begin{bmatrix}
         \Gamma_n^{-1} &  & \\
          & \ddots & \\
          &  & \Gamma_n^{-1}\\
        \end{bmatrix}, 
\end{align}
where $I_{\frac{P}{B}}$ is the identity matrix of size $\frac{P}{B}\times\frac{P}{B}$ and $P_{\alpha_n}(\ell+1)$ is a polynomial in $\alpha_n$ of degree $\ell+1$. For example, the answer of database $n$,$n\in\{1,\dots,N\}$ corresponding to the first sparse subpacket of segment 1 (i.e., $\Tilde{V}_1(1)=1$) is given by,
\begin{align}
    A_n^{[\Tilde{V}_1(1)]}&=(D_n\times S_n)^T Q_n^{[\Tilde{V}_1(1)]}\\
    &=\left(\begin{bmatrix}
        \Gamma_n^{-1} & 0_{\ell\times\ell} & 0_{\ell\times\ell} & 0_{\ell\times\ell} & 0_{\ell\times\ell}\\
        0_{\ell\times\ell} & \Gamma_n^{-1} &0_{\ell\times\ell} &  0_{\ell\times\ell} & 0_{\ell\times\ell}\\
        0_{\ell\times\ell} & 0_{\ell\times\ell} & \Gamma_n^{-1} & 0_{\ell\times\ell}  & 0_{\ell\times\ell}\\
        0_{\ell\times\ell} & 0_{\ell\times\ell} &  0_{\ell\times\ell}& \Gamma_n^{-1} & 0_{\ell\times\ell} \\
        0_{\ell\times\ell}  & 0_{\ell\times\ell} & 0_{\ell\times\ell} & 0_{\ell\times\ell}& \Gamma_n^{-1}\\
    \end{bmatrix}
    \begin{bmatrix}
        \begin{bmatrix}
    \frac{1}{f_{1}-\alpha_n}W_{1}^{[1]}+\sum_{j=0}^{\ell}\alpha_n^j I_{1,j}\\
    \vdots\\
    \frac{1}{f_{\ell}-\alpha_n}W_{\ell}^{[1]}+\sum_{j=0}^{\ell}\alpha_n^j I_{\ell,j}
    \end{bmatrix}\\
    \vdots\\
    \begin{bmatrix}
    \frac{1}{f_{1}-\alpha_n}W_{1}^{[5]}+\sum_{j=0}^{\ell}\alpha_n^j I_{1,j}\\
    \vdots\\
    \frac{1}{f_{\ell}-\alpha_n}W_{\ell}^{[5]}+\sum_{j=0}^{\ell}\alpha_n^j I_{\ell,j}
    \end{bmatrix}
    \end{bmatrix}\right)^T
    \left(\begin{bmatrix}
        0_{\ell}\\\frac{1}{f_1-\alpha_n}\\\vdots\\\frac{1}{f_\ell-\alpha_n}\\0_{\ell}\\0_{\ell}\\0_{\ell}\\
    \end{bmatrix}+Z_1\right)\\
    &=\frac{1}{f_1-\alpha_n}W_1^{[2]}+\dotsc+\frac{1}{f_\ell-\alpha_n}W_\ell^{[2]}+P_{\alpha_n}(\ell+1).\label{eg_ans}
\end{align}
Now, the users obtain the parameters of real subpacket 2 of segment 1, (i.e., $V_1(1)=\Tilde{P}_1(\Tilde{V}_1(1))=2$) by solving,
\begin{align}\label{mat_c1}
    \begin{bmatrix}        A_1^{\Tilde{V}_1(1)}\\\vdots\\A_N^{\Tilde{V}_1(1)}
    \end{bmatrix}
= 
& \begin{bmatrix}
    \frac{1}{f_1-\alpha_1} & \dotsc & \frac{1}{f_\ell-\alpha_1} & 1 & \alpha_1 & \dotsc & \alpha_1^{\ell+1}\\
    \vdots & \vdots & \vdots & \vdots & \vdots & \vdots & \vdots\\
    \frac{1}{f_1-\alpha_N} & \dotsc & \frac{1}{f_\ell-\alpha_N} & 1 & \alpha_N & \dotsc & \alpha_N^{\ell+1}\\
\end{bmatrix}
\begin{bmatrix}
        W_{1}^{[2]}\\\vdots\\W_{\ell}^{[2]}\\ R_0\\\vdots\\R_{\ell+1}
    \end{bmatrix},
\end{align}
where $R_i$ are the coefficents of the polynomial $P_{\alpha_n}(\ell+1)$ in \eqref{eg_ans}. Note that \eqref{mat_c1} (and the corresponding general set of equations in \eqref{general_r}) is solvable given that $N=2\ell+2$, which determines the subpacketization as $\ell=\frac{N-2}{2}$. The same procedure described above is carried out for all sparse subpackets in each of the $B$ segments. The resulting reading cost (including both data and permuted index downloads) is given by,
\begin{align}
    C_R&=\frac{Pr'(\log_q \frac{P}{B}+\log_q B)+Pr'N}{L}=\frac{Pr'(\log_qP+N)}{P\frac{N-2}{2}}=\frac{2r'(1+\frac{\log_q P}{N})}{1-\frac{2}{N}}.
\end{align}

\emph{Writing Phase:} In the writing phase, each user selects the $Pr$ subpackets with the most significant updates and sends the corresponding noise added combined updates (single bit per subpacket) along with their permuted subpacket indices to each of the databases. The noise added combined update of real subpacket $i$ of segment $j$ (assuming this subpacket is among the $Pr$ selected subpackets) sent to database $n$, $n\in\{1,\dotsc,N\}$) is given by,
\begin{align}\label{comb}
    U_n^{[i,j]}=\sum_{k=1}^\ell\prod_{r=1,r\neq k}^\ell (f_r-\alpha_n)\Tilde{\Delta}_{k}^{[i,j]}+\prod_{r=1}^\ell(f_r-\alpha_n)Z^{[i,j]},
\end{align}
where $\Tilde{\Delta}_{k}^{[i,j]}=\frac{\Delta_{k}^{[i,j]}}{\prod_{r=1,r\neq k}^\ell(f_r-f_k)}$ with $\Delta_{k}^{[i,j]}$ being the update of the $k$th bit of the sparse subpacket $i$ of segment $j$ and $Z^{[i,j]}$ is a random noise bit. To determine the permuted subpacket index of subpacket $i$ of segment $j$, consider the permutation assigned for segment $j$ (i.e., $\Tilde{P}_j$) to be a one-to-one mapping from the set $\{1,\dotsc,\frac{P}{B}\}$ to the set $\Tilde{P}_j$ in the exact order. Then, the permuted subpacket index corresponding to subpacket $i$ of segment $j$ is given by,
\begin{align}\label{perm_id1}
    Y^{[i,j]}=\Tilde{P}_j^{-1}(i), \quad j\in\{1,\dotsc,B\}.
\end{align}
Once the combined updates and permuted subpacket indices corresponding to all $Pr$ chosen subpackets are computed, the user uploads the $Pr$ (update, subpacket, segment) tuples to all databases.\footnote{Note that the `subpacket' and `segment' elements in the (update, subpacket, segment) refer to the permuted subpacket index and the real segment index, respectively.}

For example, assume that a given user wants to update the real subpackets 2 and 4 from segment 1, subpacket 2 from segment 2 and subpacket 5 from segment 3. Based on the permutations considered in this example, i.e., $\Tilde{P}_1=\{2,1,4,5,3\}$, $\Tilde{P}_2=\{3,5,2,4,1\}$ and $\Tilde{P}_3=\{5,2,3,1,4\}$, the user generates the combined updates $U_n^{[2,1]}$, $U_n^{[4,1]}$, $U_n^{[2,2]}$ and $U_n^{[5,3]}$ which are of the form \eqref{comb}. The corresponding permuted subpacket indices are given by,
\begin{align}
    Y^{[2,1]}&=\Tilde{P}_1^{-1}(2)=1\\
    Y^{[4,1]}&=\Tilde{P}_1^{-1}(4)=3\\
    Y^{[2,2]}&=\Tilde{P}_2^{-1}(2)=3\\
    Y^{[5,3]}&=\Tilde{P}_3^{-1}(5)=1.
\end{align}
Therefore, the two permuted (update, subpacket, segment) tuples corresponding to segment 1, sent by the user to database $n$ are given by, $(U_n^{[2,1]},1,1)$ and $(U_n^{[4,1]},3,1)$. Similarly, the permuted (update, subpacket, segment) tuples corresponding to segments 2 and 3 are given by $(U_n^{[2,2]},3,2)$ and $(U_n^{[5,3]},1,3)$, respectively.\footnote{Note that there is no permutation in the segment index, and only the subpacket indices within each segment is being permuted.} Once database $n$, $n\in\{1,\dotsc,N\}$ receives the $Pr$ (update, subpacket, segment) tuples, it creates the permuted update vectors $\hat{Y}_n^{[j]}$ for each segment $j\in\{1,\dotsc,B\}$ given by,
\begin{align}\label{perm_upd1}
    \hat{Y}_n^{[j]}=\sum_{i=1}^{\frac{P}{B}} U_n^{[i,j]}e_{\frac{P}{B}}(Y^{[i,j]}),
\end{align}
where $e_{\frac{P}{B}}(Y^{[i,j]})$ is the all zeros vector of size $\frac{P}{B}\times1$ with a $1$ at the $Y^{[i,j]}$th position. Note that we consider $U_n^{[i,j]}=0$ for those values of $i$, $i\in\{1,\dotsc,\frac{P}{B}\}$ whose corresponding subpackets are not included in the set of $Pr$ selected subpackets. In order to reverse the permutation privately, each database creates,
\begin{align}
    \hat{U}_n^{[j]}=\hat{Y}_n^{[j]}\otimes 1_{\ell}= [\hat{Y}_n^{[j]}(1)\cdot1_\ell,\dotsc,\hat{Y}_n^{[j]}(\frac{P}{B})\cdot 1_\ell]^T,
\end{align}
for each segment $j$, where $1_\ell$ is the all ones vector of size $\ell\times1$. For the example considered, the $\hat{U}_n^{[j]}$ vectors for the three segments, generated by database $n$, $n\in\{1,\dots,N\}$ based on the received information $(U_n^{[2,1]},1,1)$, $(U_n^{[4,1]},3,1)$,$(U_n^{[2,2]},3,2)$ and $(U_n^{[5,3]},1,3)$ are given  by,
\begin{align}\label{perm_upd}
    \hat{U}_n^{[1]}=\begin{bmatrix}        U_n^{[2,1]}\cdot1_\ell\\0\cdot1_\ell\\U_n^{[4,1]}\cdot1_\ell\\0\cdot1_\ell\\0\cdot1_\ell
    \end{bmatrix}, \quad
    \hat{U}_n^{[2]}=\begin{bmatrix}        0\cdot1_\ell\\0\cdot1_\ell\\U_n^{[2,2]}\cdot1_\ell\\0\cdot1_\ell\\0\cdot1_\ell
    \end{bmatrix}, \quad
    \hat{U}_n^{[3]}=\begin{bmatrix}        U_n^{[5,3]}\cdot1_\ell\\0\cdot1_\ell\\0\cdot1_\ell\\0\cdot1_\ell\\0\cdot1_\ell
    \end{bmatrix}.
\end{align}
Next, the databases privately rearrange the updates in the real order and calculate the incremental updates of each segment as $\bar{U}_n^{[j]}=R_n^{[j]}\hat{U}_n^{[j]}$ for $j\in\{1,\dotsc,B\}$, and add it to the $j$th segment of the existing storage to obtain the updated storage. Consider the incremental update calculation of segment 1 in database $n$, $n\in\{1,\dots,N\}$ for the example considered,
\begin{align}
    \bar{U}_n^{[1]}&=R_n^{[1]}\hat{U}_n^{[1]}\\
    &=\left(\begin{bmatrix}
        0_{\ell\times\ell} & \Gamma_n & 0_{\ell\times\ell} & 0_{\ell\times\ell} & 0_{\ell\times\ell}\\
        \Gamma_n & 0_{\ell\times\ell} & 0_{\ell\times\ell} & 0_{\ell\times\ell} & 0_{\ell\times\ell}\\
        0_{\ell\times\ell} & 0_{\ell\times\ell} & 0_{\ell\times\ell}  & 0_{\ell\times\ell}& \Gamma_n\\
        0_{\ell\times\ell} & 0_{\ell\times\ell} & \Gamma_n & 0_{\ell\times\ell}&  0_{\ell\times\ell} \\
        0_{\ell\times\ell}  & 0_{\ell\times\ell} & 0_{\ell\times\ell} & \Gamma_n & 0_{\ell\times\ell}\\
    \end{bmatrix}+\bar{Z}_1\right)\begin{bmatrix}        U_n^{[2,1]}\cdot1_\ell\\0\cdot1_\ell\\U_n^{[4,1]}\cdot1_\ell\\0\cdot1_\ell\\0\cdot1_\ell
    \end{bmatrix}\\
    &=\begin{bmatrix}
        0_\ell\\\frac{U_n^{[2,1]}}{f_1-\alpha_n}\\\vdots\\\frac{U_n^{[2,1]}}{f_\ell-\alpha_n}\\0_\ell\\\frac{U_n^{[4,1]}}{f_1-\alpha_n}\\\vdots\\\frac{U_n^{[4,1]}}{f_\ell-\alpha_n}\\0_\ell
    \end{bmatrix}+P_{\alpha_n}(\ell)=\begin{bmatrix}
        0_\ell\\\frac{\Delta_{1}^{[2,1]}}{f_1-\alpha_n}\\\vdots\\\frac{\Delta_{\ell}^{[2,1]}}{f_\ell-\alpha_n}\\0_\ell\\\frac{\Delta_{1}^{[4,1]}}{f_1-\alpha_n}\\\vdots\\\frac{\Delta_{\ell}^{[4,1]}}{f_\ell-\alpha_n}\\0_\ell
    \end{bmatrix}+P_{\alpha_n}(\ell)\label{incr_c1},
\end{align}
where $P_{\alpha_n}(\ell)$ here is a vector of size $\frac{P\ell}{B}$ consisting of polynomial in $\alpha_n$ of degree $\ell$, and the last equality is obtained by applying Lemma 1 in \cite{pruw_jpurnal}. The same process is carried out for the other two segments as well. Since the incremental update is in the same form as the storage in \eqref{storage_c1_eg}, the storage of segment $j$, $j\in\{1,2,3\}$ at time $t$ can be updated as,
\begin{align}
    S_n^{[j]}(t)=S_n^{[j]}(t-1)+\bar{U}_n^{[j]}, \quad n\in\{1,\dots,N\}.
\end{align}
Note from \eqref{incr_c1} that for segment 1, the two real sparse subpackets 2 and 4 have been correctly updated, while ensuring that the rest of the subpackets remain the same, without revealing the real subpacket indices 2 and 4 to any of the databases. The resulting writing cost is given by,
\begin{align}
    C_W=\frac{PrN(1+\log_q B+\log_q \frac{P}{B})}{L}=\frac{PrN(1+\log_q P)}{P\frac{N-2}{2}}=\frac{2r(1+\log_qP)}{1-\frac{2}{N}}.
\end{align}
The total storage complexity (including both data and the permutation reversing matrices) is given by $O(L)+O(\frac{L^2}{B^2}\times B)=O(\frac{L^2}{B})$. The information leakage is derived in Section~\ref{info_leak}.

\textbf{Case 2:} MDS coded storage and smaller permutation reversing matrices are used in this case to reduce the storage cost, at the expense of a larger communication cost. The information leakage is the same for both cases 1 and 2, since they both use only within-segment permutations. The same example considered for case 1 (shown in Fig.~\ref{init_c1_eg}) is considered in this case as well.

\emph{Initialization:} A single subpacket $s$ in case 2 is stored in database $n$, $n\in\{1,\dotsc,N\}$ as,
\begin{align}
    S_n^{[s]}=\sum_{i=1}^\ell \frac{1}{\alpha_n^i}W_i^{[s]}+\sum_{i=0}^\ell \alpha_n^i Z_{i}^{[s]}.
\end{align}
Therefore, the storage of segment $j$, $j\in\{1,\dotsc,B\}$ is given by,
 \begin{align}\label{storage_c2_eg}
 S_n^{[j]}=\begin{bmatrix}
    \sum_{i=1}^\ell \frac{1}{\alpha_n^i}W_i^{[1,j]}+\sum_{i=0}^\ell \alpha_n^i Z_{i}^{[1,j]}\\
    \vdots\\
    \sum_{i=1}^\ell \frac{1}{\alpha_n^i}W_i^{[\frac{P}{B},j]}+\sum_{i=0}^\ell \alpha_n^i Z_{i}^{[\frac{P}{B},j]}
    \end{bmatrix},
\end{align}
where $W_i^{[s,j]}$ is the $i$th parameter of subpacket $s$ in segment $j$ and $Z^{[s,j]}_i$ are random noise symbols. Note that $\frac{P}{B}=5$ for the example considered. Similar to case 1, the coordinator initializes all noise terms in storage, assigns $B$ permutations $\tilde{P}_i$, $i\in\{1,\dotsc,B\}$ of the subpackets in each of the $B$ segments and sends them to the users, and sends the corresponding $B$ noise added permutation reversing matrices $R_n^{[i]}$, $i\in\{1,\dotsc,B\}$ to database $n$, $n\in\{1,\dotsc.N\}$, as shown in Fig.~\ref{init_c1_eg}. The noise added permutation reversing matrices are of the form,
\begin{align}\label{reverse2}    R_n^{[i]}=\bar{R}^{[i]}+\alpha_n^{\ell}\bar{Z}^{[i]}, \quad i\in\{1,\dotsc,B\},
\end{align}
where $\bar{R}^{[i]}$ is the permutation reversing matrix corresponding to the $i$th permutation $\tilde{P}_i$, and $\bar{Z}^{[i]}$ is a random noise matrix, both of size $\frac{P}{B}\times\frac{P}{B}$. The noise added permutation reversing matrices corresponding to the three segments, sent to database $n$, $n\in\{1,\dots,N\}$ for the example in Fig.~\ref{init_c1_eg} are given by (recall: $\Tilde{P}_1=(2,1,4,5,3)$, $\Tilde{P}_2=(3,5,2,4,1)$, $\Tilde{P}_3=(5,2,3,1,4)$),
\begin{align}\label{R1_c2}
    R_n^{[1]}=\!\!\begin{bmatrix}
        0 & 1 & 0 & 0 & 0\\
        1 & 0 & 0 & 0 & 0\\
        0 & 0 & 0 & 0 & 1\\
        0 & 0 & 1 & 0 & 0\\
        0  & 0 & 0 & 1 & 0\\
    \end{bmatrix}\!\!+\alpha_n^\ell\bar{Z}^{[1]},\  R_n^{[2]}=\!\!\begin{bmatrix}
        0 & 0 & 0 & 0 & 1\\
        0 & 0 & 1 & 0 & 0\\
        1 & 0 & 0 & 0 & 0\\
        0 & 0 & 0 & 1 & 0\\
        0  & 1 & 0 & 0 & 0\\
    \end{bmatrix}\!\!+\alpha_n^\ell\bar{Z}^{[2]},\  R_n^{[3]}=\!\!\begin{bmatrix}
        0 & 0 & 0 & 1 & 0\\
        0 & 1 & 0 & 0 & 0\\
        0 & 0 & 1 & 0 & 0\\
        0 & 0 & 0 & 0 & 1\\
        1  & 0 & 0 & 0 & 0\\
    \end{bmatrix}\!\!+\alpha_n^\ell\bar{Z}^{[3]}.
\end{align}

As explained in case 1, the coordinator leaves the system once the FL process begins, and all subsequent communications take place only between the users and databases using permuted subpacket indices.

\emph{Reading Phase:} As described in case 1, let $\tilde{V}_j$ be the set of permuted indices of the sparse subpackets chosen from segment $j$ to be sent to the users for $j\in\{1,\dotsc,B\}$. For example, let $\Tilde{V}_1=\{1,3\}$ be the permuted set of sparse subpackets of segment 1 that needs to be sent to the users at time $t$. One designated database sends the permuted subpacket indices of each segment (segment 1: $\Tilde{V}_1=\{1,3\}$) to the users, from which the users identify the corresponding real sparse subpacket indices using the known permutations using $V_j(i)=\tilde{P}_j(\tilde{V}_j(i))$, where $V_j$ is the vector containing the real indices of the sparse subpackets in segment $j$. In particular, the users perform the same calculation in \eqref{real_id} for segment 1 as well as for the other two segments, based on the received sets $\Tilde{V}_2$ and $\Tilde{V}_3$. Similar to case 1, each database generates a query to send each of the chosen subpackets. The query corresponding to the $i$th permuted sparse subpacket of segment $j$ is given by,
\begin{align}
    Q_n^{[\Tilde{V}_j(i)]}=R_n^{[j]}(:,\Tilde{V}_j(i)), \quad j\in\{1,\dotsc,B\}.
\end{align}
Following are the two queries generated by database $n$, $n\in\{1,\dots,N\}$ to send the two sparse subpackets (with permuted indices $\Tilde{V}_1=\{1,3\}$) of segment 1 to the users,
\begin{align}
    Q_n^{[\Tilde{V}_1(1)]}=Q_n^{[1]}=R_n^{[1]}(:,1)=\begin{bmatrix}
        0\\1\\0\\0\\0
    \end{bmatrix}+\alpha_n^\ell\hat{Z}_1, \quad \text{and} \quad Q_n^{[\Tilde{V}_1(2)]}=Q_n^{[3]}=R_n^{[1]}(:,3)=\begin{bmatrix}
        0\\0\\0\\1\\0
    \end{bmatrix}+\alpha_n^\ell\hat{Z}_3,
\end{align}
where $\hat{Z}_1$ and $\hat{Z}_3$ are the first and third columns of $\bar{Z}^{[1]}$ in \eqref{R1_c2}. Then, database $n$, $n\in\{1,\dotsc,N\}$ sends the answers corresponding to each sparse subpacket of each segment to the users as,
\begin{align}
    A_n^{[\tilde{V}_j(i)]}&=(S_n^{[j]})^TQ_n^{[\tilde{V}_j(i)]},\quad j\in\{1,\dotsc,B\}\\
    &=\sum_{i=1}^\ell\frac{1}{\alpha_n^i}W_i^{[V_j(i),j]}+P_{\alpha_n}(2\ell),\label{ans2}
\end{align}
where $P_{\alpha_n}(2\ell)$ is a polynomial in $\alpha_n$ of degree $2\ell$. For example, the answer corresponding to the first sparse subpacket of segment 1 ($\Tilde{V}_1(1)=1$), sent by database $n$, $n\in\{1,\dots,N\}$ to the users is given by,
\begin{align}
   A_n^{[\Tilde{V}_1(1)]}&=(S_n^{[1]})^TQ_n^{[\Tilde{V}_1(1)]}\\
    &=\begin{bmatrix}
    \sum_{i=1}^\ell \frac{1}{\alpha_n^i}W_i^{[1,1]}+\sum_{i=0}^\ell \alpha_n^i Z^{[1,1]}_i\\
    \vdots\\
    \sum_{i=1}^\ell \frac{1}{\alpha_n^i}W_i^{[5,1]}+\sum_{i=0}^\ell \alpha_n^i Z^{[5,1]}_i
    \end{bmatrix}^T\left(\begin{bmatrix}
        0\\1\\0\\0\\0
    \end{bmatrix}+\alpha_n^\ell\hat{Z}_1\right)\\
    &=\sum_{i=1}^\ell\frac{1}{\alpha_n^i}W_i^{[2,1]}+P_{\alpha_n}(2\ell).\label{eg2}
\end{align}
The users obtain the parameters of the real subpacket 2 of segment 1, (i.e., $V_1(1)=\Tilde{P}_1(\Tilde{V}_1(1))=2$) by solving,
\begin{align}\label{mat_c2}
    \begin{bmatrix}
        A_1^{\Tilde{V}_1(1)}\\\vdots\\A_N^{\Tilde{V}_1(1)}
    \end{bmatrix}
= 
& \begin{bmatrix}
    \frac{1}{\alpha_1^\ell} & \dotsc & \frac{1}{\alpha_1} & 1 & \alpha_1 & \dotsc & \alpha_1^{2\ell}\\
    \vdots & \vdots & \vdots & \vdots & \vdots & \vdots & \vdots\\
    \frac{1}{\alpha_N^\ell} & \dotsc & \frac{1}{\alpha_N} & 1 & \alpha_N & \dotsc & \alpha_N^{2\ell}\\
\end{bmatrix}
\begin{bmatrix}
        W_{\ell}^{[2,1]}\\\vdots\\W_{1}^{[2,1]}\\ R_0\\\vdots\\R_{2\ell}
    \end{bmatrix},
\end{align}
where $R_i$ are the coefficients of the polynomial in \eqref{eg2}. Note that \eqref{mat_c2} (and also the general answers in \eqref{ans2}) is solvable given that $N=3\ell+1$, which determines the subpacketization as $\ell=\frac{N-1}{3}$. The same procedure described above is carried out for all sparse subpackets in each of the $B$ segments. The resulting reading cost is given by,
\begin{align}
    C_R&=\frac{Pr'\log_q P+Pr'N}{L}=\frac{Pr'(\log_qP+N)}{P\frac{N-1}{3}}=\frac{3r'(1+\frac{\log_q P}{N})}{1-\frac{1}{N}}.
\end{align}

\emph{Writing Phase:} In the writing phase, the user generates $Pr$ combined updates corresponding to the $Pr$ subpackets with the most significant updates. The combined update of the $i$th subpacket of segment $j$ is defined as (assuming this subpacket is among the $Pr$ selected subpackets),
\begin{align}
    U_n^{[i,j]}=\sum_{k=1}^\ell \frac{1}{\alpha_n^k}\Delta_{k}^{[i,j]}+Z^{[i,j]},
\end{align}
where $\Delta_{k}^{[i,j]}$ is the update of the $k$th bit of the $i$th subpacket of segment $j$ and $Z^{[i,j]}$ is a random noise symbol. Similar to case 1, the user generates the permuted subpacket indices corresponding to the real subpacket indices $i$ of each segment $j$, of the selected $Pr$ subpackets, using \eqref{perm_id1}. Once the permuted subpacket indices are generated, the user sends the permuted (update, subpacket, segment) tuples of the $Pr$ selected subpackets to all databases similar to case 1. For the same example considered in case 1 where the user wants to update the real subpackets 2 and 4 from segment 1, subpacket 2 from segment 2 and subpacket 5 from segment 3, the user sends the same permuted (update, subpacket, segment) tuples sent in case 1 given by, $(U_n^{[2,1]},1,1), (U_n^{[4,1]},3,1)$, $(U_n^{[2,2]},3,2)$, $(U_n^{[5,3]},1,3)$ to database $n$, $n\in\{1,\dotsc,N\}$ assuming the same three permutations given by $\Tilde{P}_1=\{2,1,4,5,3\}$, $\Tilde{P}_2=\{3,5,2,4,1\}$ and $\Tilde{P}_3=\{5,2,3,1,4\}$, for the three segments. Once the databases receive the $Pr$ (update, subpacket, segment) tuples, they create the permuted update vectors $\hat{Y}_n^{[j]}$ for each segment $j\in\{1,\dotsc,B\}$ using \eqref{perm_upd1}. For the example considered, the three permuted update vectors created at database $n$ are given by,
\begin{align}
    \hat{Y}_n^{[1]}=\begin{bmatrix}        U_n^{[2,1]}\\0\\U_n^{[4,1]}\\0\\0
    \end{bmatrix},\quad
    \hat{Y}_n^{[2]}=\begin{bmatrix}        0\\0\\U_n^{[2,2]}\\0\\0
    \end{bmatrix},\quad
    \hat{Y}_n^{[3]}=\begin{bmatrix}        U_n^{[5,3]}\\0\\0\\0\\0
    \end{bmatrix},
\end{align}
based on the received information $(U_n^{[2,1]},1,1), (U_n^{[4,1]},3,1)$, $(U_n^{[2,2]},3,2)$, $(U_n^{[5,3]},1,3)$. Using the permuted update vectors $\hat{Y}_n^{[j]}$, $j\in\{1,\dotsc,B\}$, database $n$, $n\in\{1,\dotsc,N\}$ privately calculates the correctly rearranged incremental update vector of each segment as $\bar{U}_n^{[j]}=R_n^{[j]}\hat{Y}_n^{[j]}$, $j\in\{1,\dotsc,B\}$. The resulting incremental update is of the same form as the storage in \eqref{storage_c2_eg}, and therefore can be added to the existing storage to obtain the updated storage of each segment. To explain the above process in terms of an example, consider the incremental update of segment 1 in the same example considered so far,
\begin{align}
    \bar{U}_n^{[1]}&=R_n^{[1]}\hat{Y}_n^{[1]}\\
    &=\left(\begin{bmatrix}
        0 & 1 & 0 & 0 & 0\\
        1 & 0 & 0 & 0 & 0\\
        0 & 0 & 0 & 0 & 1\\
        0 & 0 & 1 & 0 & 0\\
        0  & 0 & 0 & 1 & 0\\    \end{bmatrix}+\alpha_n^\ell\bar{Z}^{[1]}\right)
        \begin{bmatrix}U_n^{[2,1]}\\0\\U_n^{[4,1]}\\0\\0
    \end{bmatrix}\\
    &=\begin{bmatrix}
        0\\U_n^{[2,1]}\\0\\U_n^{[4,1]}\\0
    \end{bmatrix}+P_{\alpha_n}(\ell)=\begin{bmatrix}
        0\\\sum_{i=1}^\ell \frac{1}{\alpha_n^i}\Delta_{i}^{[2,1]}\\0\\\sum_{i=1}^\ell \frac{1}{\alpha_n^i}\Delta_{i}^{[4,1]}\\0
    \end{bmatrix}+P_{\alpha_n}(\ell)\label{incr_c2},
\end{align}
where $P_{\alpha_n}(\ell)$ here is a vector of size $5\times1$, consisting of polynomials in $\alpha_n$ of degree $\ell$. Since the incremental update of each segment (i.e., \eqref{incr_c2}) is in the same form as the storage in \eqref{storage_c2_eg}, the incremental update is directly added to the existing storage to obtain it's updated version, i.e.,
\begin{align}
    S_n^{[j]}(t)=S_n^{[j]}(t-1)+\bar{U}_n^{[j]},\quad j\in\{1,\dotsc,B\},\ n\in\{1,\dots,N\}.
\end{align}
The writing cost for case 2 is given by,
\begin{align}
    C_W=\frac{PrN(1+\log_q B+\log_q \frac{P}{B})}{L}=\frac{PrN(1+\log_q P)}{P\frac{N-1}{3}}=\frac{3r(1+\log_qP)}{1-\frac{1}{N}}.
\end{align}
The total storage complexity (including both data and the permutation reversing matrices) is given by $O(P)+O(\frac{P^2}{B^2}\times B)=O(\frac{P^2}{B})=O(\frac{L^2}{BN^2})$. The information leakage is derived in Section~\ref{info_leak}.

\textbf{Case 3:} In this case, we use uncoded storage with large permutation reversing matrices. Note that in both cases 1 and 2, only the subpacket indices within each segment were permuted, and the real segment indices were uploaded to the databases by the users. In this case, we permute subpacket indices within segments as well as the segment indices to reduce the information leakage further. However, this increases the storage cost since the permutation of segment indices requires an additional noise added permutation reversing matrix to be stored at the databases. The communication cost is not significantly affected by the additional round of permutation, compared to case 1 (lowest communication cost thus far).

\begin{figure*}
		\centering
		\includegraphics[scale=1]{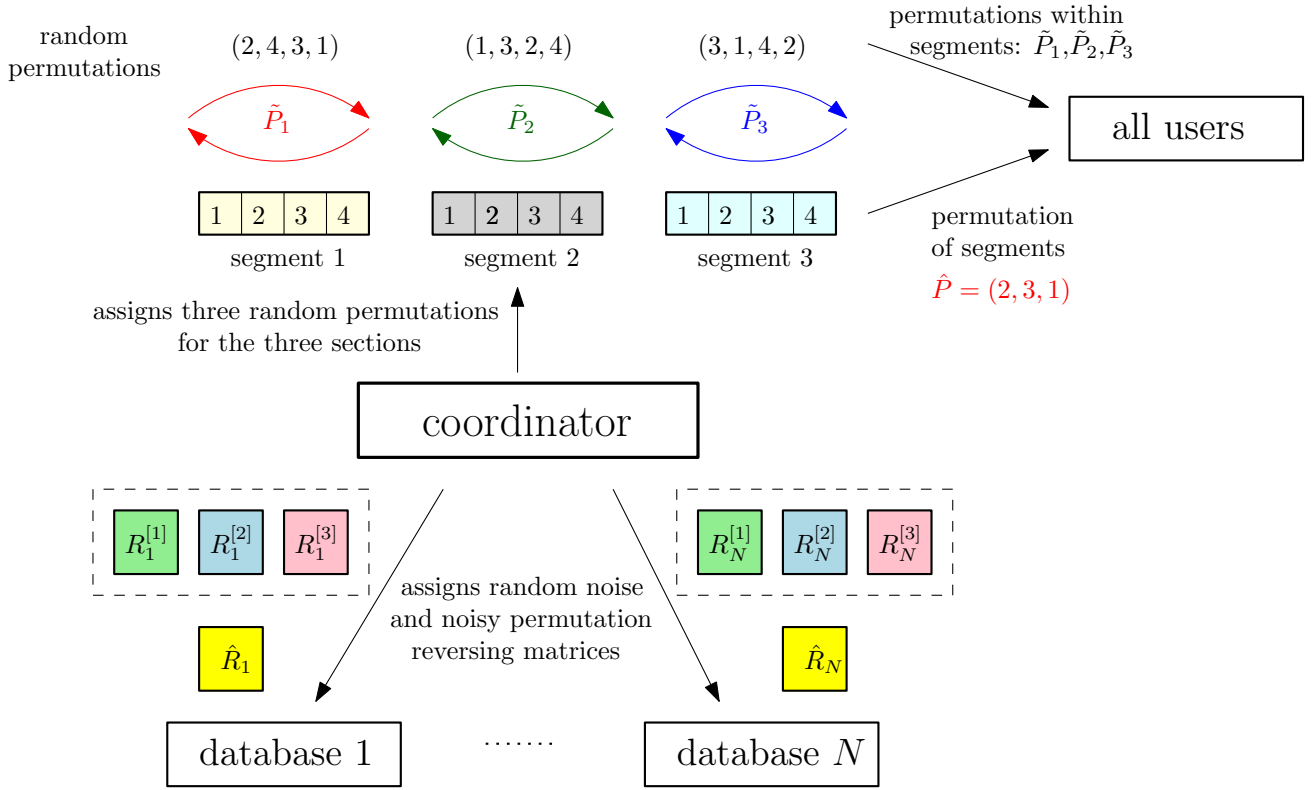}
		\caption{Initialization of the scheme for cases 3 and 4.}
		\label{init_c3_eg}
\end{figure*}

For cases 3 and 4, we present the general scheme along with the example setting with $P=12$ subpackets (with subpacketization $\ell$) which are divided into and $B=3$ equal segments, as shown in Fig.~\ref{init_c3_eg}.

\emph{Initialization:} The storage of a single subpacket (subpacket $s$) in case 3 is given by,
 \begin{align}\label{storage_c3_eg}
 S_n^{[s]}=\begin{bmatrix}
    \frac{1}{f_{1}-\alpha_n}W_{1}^{[s]}+\sum_{j=0}^{\ell+1}\alpha_n^j Z_{1,j}^{[s]}\\
    \vdots\\
    \frac{1}{f_{\ell}-\alpha_n}W_{\ell}^{[s]}+\sum_{j=0}^{\ell+1}\alpha_n^j Z_{\ell,j}^{[s]}
    \end{bmatrix},
\end{align}
with the same notation as in case 1. The subpackets are stacked one after the other (subpacket $1$ through subpacket $\frac{P}{B}$ in each segment) in the order of segment $1$ through segment $B$. At the initialization stage, the coordinator sends $B$ randomly and independently chosen permutations of the $\frac{P}{B}$ subpackets in each of the $B$ segments , $\tilde{P}_1$, $\dotsc$, $\tilde{P}_B$, as well as a randomly and independently chosen permutation of the $B$ segments $\hat{P}$ to the users. The coordinator also places the corresponding noise added permutation reversing matrices (corresponding to $B$ within-segments permutations: $R_n^{[1]}$,$\dotsc$,$R_n^{[B]}$ and one inter-segment permutation: $\hat{R}_n$) at each database $n$, $n\in\{1,\dotsc,N\}$. The noise added permutation reversing matrix corresponding to the $i$th segment, $i\in\{1,\dotsc,B\}$ stored in database $n$, $n\in\{1,\dotsc,N\}$ is given by,
\begin{align}\label{R_i}
    R_n^{[i]}=(\tilde{R}^{[i]}\otimes\Gamma_n)+\Tilde{Z}^{[i]},
\end{align}
with the same notation as in case 1. The noise added permutation reversing matrix corresponding to the inter-segment permutation $\hat{P}$ stored in database $n$, $n\in\{1,\dotsc,N\}$ is given by,
\begin{align}\label{R_hat}
    \hat{R}_n=(\bar{R}\otimes I_{\ell})+(I_B\otimes\Gamma_n^{-1})\hat{Z}=\begin{bmatrix}
        b^{[n]}_{1,1}&\dotsc&b^{[n]}_{1,B}\\
        \vdots&\ddots&\vdots\\
        b^{[n]}_{B,1}&\dotsc&b^{[n]}_{B,B}
    \end{bmatrix},
\end{align}
where $\bar{R}$ is the permutation reversing matrix corresponding to the inter-segment permutation $\hat{P}$, $I_{k}$ is the identity matrix of size $k\times k$, $\Gamma_n^{-1}$ is the diagonal matrix given by,
\begin{align}
   \Gamma_n^{-1}=\begin{bmatrix}
    f_1-\alpha_n &  & \\
    & \ddots & \\
    & &  f_\ell-\alpha_n\\
\end{bmatrix} 
\end{align}
and $\hat{Z}$ is a random noise matrix of size $B\ell\times B\ell$. Each matrix $\hat{R}_n$ is represented in blocks of size $\ell\times\ell$, as shown in the last equality in \eqref{R_hat}. According to the given permutations in Figure~\ref{init_c3_eg}, the noise added permutation reversing matrix corresponding to the first within-segment permutation ($\Tilde{P}_1=(2,4,3,1)$), i.e., $R_n^{[1]}$ is given by,
\begin{align}\label{R1}
    R_n^{[1]}=\left(\begin{bmatrix}
        0 & 0 & 0 & 1\\
        1 & 0 & 0 & 0\\
        0 & 0 & 1 & 0\\
        0 & 1 & 0 & 0
    \end{bmatrix}\otimes\Gamma_n\right)+\Tilde{Z}^{[1]}=
    \begin{bmatrix}
        0_{\ell\times\ell} & 0_{\ell\times\ell} & 0_{\ell\times\ell} & \Gamma_n\\
        \Gamma_n & 0_{\ell\times\ell} & 0_{\ell\times\ell} & 0_{\ell\times\ell} \\
        0_{\ell\times\ell} & 0_{\ell\times\ell} & \Gamma_n &0_{\ell\times\ell}\\
        0_{\ell\times\ell} & \Gamma_n & 0_{\ell\times\ell}&  0_{\ell\times\ell} 
    \end{bmatrix}+\tilde{Z}^{[1]}.
\end{align}

Similarly, $R_n^{[2]}$ and $R_n^{[3]}$ are given by,
\begin{align}
    R_n^{[2]}=\begin{bmatrix}
        \Gamma_n & 0_{\ell\times\ell} & 0_{\ell\times\ell} & 0_{\ell\times\ell} \\
        0_{\ell\times\ell} & 0_{\ell\times\ell} & \Gamma_n &0_{\ell\times\ell} \\
        0_{\ell\times\ell} & \Gamma_n & 0_{\ell\times\ell}  & 0_{\ell\times\ell}\\
        0_{\ell\times\ell}  & 0_{\ell\times\ell}&  0_{\ell\times\ell} & \Gamma_n
    \end{bmatrix}+\Tilde{Z}^{[2]}, \quad R_n^{[3]}=\begin{bmatrix}
        0_{\ell\times\ell}  & \Gamma_n & 0_{\ell\times\ell} & 0_{\ell\times\ell} \\
        0_{\ell\times\ell} & 0_{\ell\times\ell} & 0_{\ell\times\ell} & \Gamma_n  \\
        \Gamma_n & 0_{\ell\times\ell} & 0_{\ell\times\ell} & 0_{\ell\times\ell}\\
        0_{\ell\times\ell} &  0_{\ell\times\ell}&  \Gamma_n &0_{\ell\times\ell} 
    \end{bmatrix}+\Tilde{Z}^{[3]},
\end{align}
        For the same example, the inter-segment noise added permutation reversing matrix for database $n$, $n\in\{1,\dotsc,N\}$ is given by,
        \begin{align}
    \hat{R}_n&=\left(\begin{bmatrix}
        0 & 0 & 1\\
        1 & 0 & 0\\
        0 & 1 & 0
    \end{bmatrix}\otimes\Phi\right)+\left(\begin{bmatrix}
        1 & 0 & 0\\
        0 & 1 & 0\\
        0 & 0 & 1
    \end{bmatrix}\otimes\Gamma_n^{-1}\right)\hat{Z}\\
    &=\begin{bmatrix}
        0_{\ell\times\ell} & 0_{\ell\times\ell} & \Phi\\
        \Phi & 0_{\ell\times\ell} & 0_{\ell\times\ell}\\
        0_{\ell\times\ell} & \Phi & 0_{\ell\times\ell}\\ 
    \end{bmatrix}+\begin{bmatrix}
        \Gamma_n^{-1} & 0_{\ell\times\ell} & 0_{\ell\times\ell}\\
        0_{\ell\times\ell} & \Gamma_n^{-1} & 0_{\ell\times\ell}\\
        0_{\ell\times\ell} & 0_{\ell\times\ell} & \Gamma_n^{-1}\\ 
    \end{bmatrix}\hat{Z}=\begin{bmatrix}
        b_{1,1}^{[n]} & b_{1,2}^{[n]} & b_{1,3}^{[n]}\\
        b_{2,1}^{[n]} & b_{2,2}^{[n]} & b_{2,3}^{[n]}\\
        b_{3,1}^{[n]} & b_{3,2}^{[n]} & b_{3,3}^{[n]}
    \end{bmatrix},\label{eg_blk}
\end{align}
where $\Phi=I_{\ell}$. Each matrix $\hat{R}_n$ is represented in blocks of size $\ell\times\ell$, as shown in the last equality in \eqref{eg_blk}, which is useful in the subsequent calculations. The coordinator leaves the system once the storage and permutations are initialized, and the noise added permutation reversing matrices are placed at the databases, before the FL process begins. All communications in the FL process are carried out between the users and databases using permuted subpacket and segment indices.

\emph{Reading Phase:} The databases determine the set of $Pr'$ subpackets to be sent to the users at time $t$, based on the permuted information received from the users in the writing phase at time $t-1$. For example, the databases consider the permuted indices of the subpackets updated by all users at time $t-1$, and select the most popular $Pr'$ of them (in terms of permuted indices) to be sent to the users in the reading phase of time $t$. In case 3, the databases are unaware of the real indices of subpackets within each segment, as well as the corresponding real inidces of the segments, updated by the users. However, the databases can communicate the sparse subpacket and segment indices with the users in their permuted versions (same as what was received in the writing phase at time $t-1$). The users are able to convert the permuted indices into their real versions as all permutations are known by the users. Let the permuted (subpacket, segment) information of each of the $Pr'$ subpackets be indicated by $(\eta_p,\phi_p)$. This information is sent to all users at time $t$ by one designated database. The users can convert each permuted $(\eta_p,\phi_p)$ into its real versions $(\eta_r,\phi_r)$ using,
\begin{align}
    \phi_r&=\hat{P}(\phi_p)\label{decode_seg}\\
    \eta_r&=\Tilde{P}_{\phi_r}(\eta_p)\label{decode_sub}.
\end{align}
For the example in Fig.~\ref{init_c3_eg}, assume that the following permuted  (subpacket, segment) pairs are received by a given user from the designated database,
\begin{align}\label{eg_perm}
    (\eta_p,\phi_p)=\{(1,3),(1,1),(1,2)\}.
\end{align}
Recall that the permutations considered in this example are given by $\Tilde{P}_1=(2,4,3,1)$, $\Tilde{P}_2=(1,3,2,4)$, $\Tilde{P}_3=(3,1,4,2)$ and $\hat{P}=(2,3,1)$. In order to see how the real subpacket and segment indices are obtained, consider the first permuted pair $(1,3)$. Since the permuted segment index is $\phi_p=3$, the corresponding real segment index is $\phi_r=\hat{P}(3)=1$. Then, the user can decode the subpacket index within the first segment as, $\eta_r=\Tilde{P}_1(1)=2$. Therefore, the real (subpacket, segment) pair corresponding to the permuted (subpacket, segment) pair $(\eta_p,\phi_p)=(1,3)$ is given by $(\eta_r,\phi_r)=(2,1)$. Similarly, the real set of (subpacket, segment) pairs corresponding to the three permuted pairs are given by,
\begin{align}\label{eg_real}
    (\eta_r,\phi_r)=\{(2,1),(1,2),(3,3)\}.
\end{align}
Once the real indices of the sparse subpackets and segments are obtained, the user downloads the corresponding subpackets, one by one. To perform the calculations in the reading phase, each database first generates a combined noise added permutation reversing matrix, that combines the within-segment and inter-segment noise added permutation reversing matrices into a single noise added permutation reversing matrix, to facilitate the subsequent calculations. The combined noise added permutation reversing matrix of database $n$, $n\in\{1,\dots,N\}$ is given by,
\begin{align}
    R_n&=\begin{bmatrix}
        R_n^{[1]} & & \\
         & \ddots & \\
         & & R_n^{[B]}
    \end{bmatrix}\times \begin{bmatrix}
        I_{\frac{P}{B}}\otimes b_{1,1}^{[n]} & \dotsc & I_{\frac{P}{B}}\otimes b_{1,B}^{[n]}\\
        \vdots &\vdots & \vdots\\
        I_{\frac{P}{B}}\otimes b_{B,1}^{[n]} & \dotsc & I_{\frac{P}{B}}\otimes b_{B,B}^{[n]}
    \end{bmatrix}\\
    &=\begin{bmatrix}
        R_n^{[1]}\begin{bmatrix}
            b^{[n]}_{1,1} &   & \\
            &\ddots &\\
            &  & b^{[n]}_{1,1}
        \end{bmatrix}_{\frac{P\ell}{B}\times\frac{P\ell}{B}}
        \dotsc \qquad
        R_n^{[1]}\begin{bmatrix}
            b^{[n]}_{1,B} &   & \\
            &\ddots &\\
            &  & b^{[n]}_{1,B}
        \end{bmatrix}_{\frac{P\ell}{B}\times\frac{P\ell}{B}}\\
        \qquad\qquad \ddots \qquad\qquad\qquad \\
        R_n^{[B]}\begin{bmatrix}
            b^{[n]}_{B,1} &   & \\
            &\ddots &\\
            &  & b^{[n]}_{B,1}
        \end{bmatrix}_{\frac{P\ell}{B}\times\frac{P\ell}{B}}
        \dotsc \qquad
        R_n^{[B]}\begin{bmatrix}
            b^{[n]}_{B,B} &   & \\
            &\ddots &\\
            &  & b^{[n]}_{B,B}
        \end{bmatrix}_{\frac{P\ell}{B}\times\frac{P\ell}{B}}
    \end{bmatrix}\\   
    &=\dot{R}_n+P_{\alpha_n}(1)\label{common},
\end{align}
where $\dot{R}_n$ is the combined permutation reversing matrix obtained by replacing the $1$ in the $i$th row of $\bar{R}$ in \eqref{R_hat} by $R^{[i]}\otimes\Gamma_n$ in \eqref{R_i}, and the zeros by all zeros matrices of size $\frac{P\ell}{B}\times\frac{P\ell}{B}$. $P_{\alpha_n}(1)$ is a matrix of size $L\times L$ consisting of elements that are degree $1$ polynomials in $\alpha_n$.

As an example, consider the generation of the combined noise added permutation reversing matrix of database $n$, $n\in\{1,\dotsc,N\}$, for the example setting in Fig.~\ref{init_c3_eg},
\begin{align}
    R_n&=\begin{bmatrix}
        R_n^{[1]} & 0 & 0\\
        0 & R_n^{[2]} & 0\\
        0 & 0 & R_n^{[3]}
    \end{bmatrix}\times \begin{bmatrix}
        I_{4}\otimes b_{1,1}^{[n]} & \dotsc & I_{4}\otimes b_{1,3}^{[n]}\\
        \vdots &\vdots & \vdots\\
        I_{4}\otimes b_{3,1}^{[n]} & \dotsc & I_{4}\otimes b_{3,3}^{[n]}
    \end{bmatrix}\\
    &=\begin{bmatrix}
        R_n^{[1]}\begin{bmatrix}
            b^{[n]}_{1,1} &   & \\
            &\ddots &\\
            &  & b^{[n]}_{1,1}
        \end{bmatrix}_{4\ell\times4\ell} R_n^{[1]}\begin{bmatrix}
            b^{[n]}_{1,2} &   & \\
            &\ddots &\\
            &  & b^{[n]}_{1,2}
        \end{bmatrix}_{4\ell\times4\ell} R_n^{[1]}\begin{bmatrix}
            b^{[n]}_{1,3} &   & \\
            &\ddots &\\
            &  & b^{[n]}_{1,3}
        \end{bmatrix}_{4\ell\times4\ell}\\
        R_n^{[2]}\begin{bmatrix}
            b^{[n]}_{2,1} &   & \\
            &\ddots &\\
            &  & b^{[n]}_{2,1}
        \end{bmatrix}_{4\ell\times4\ell} R_n^{[2]}\begin{bmatrix}
            b^{[n]}_{2,2} &   & \\
            &\ddots &\\
            &  & b^{[n]}_{2,2}
        \end{bmatrix}_{4\ell\times4\ell} R_n^{[2]}\begin{bmatrix}
            b^{[n]}_{2,3} &   & \\
            &\ddots &\\
            &  & b^{[n]}_{2,3}
        \end{bmatrix}_{4\ell\times4\ell}\\
        R_n^{[3]}\begin{bmatrix}
            b^{[n]}_{3,1} &   & \\
            &\ddots &\\
            &  & b^{[n]}_{3,1}
        \end{bmatrix}_{4\ell\times4\ell} R_n^{[3]}\begin{bmatrix}
            b^{[n]}_{3,2} &   & \\
            &\ddots &\\
            &  & b^{[n]}_{3,2}
        \end{bmatrix}_{4\ell\times4\ell} R_n^{[3]}\begin{bmatrix}
            b^{[n]}_{3,3} &   & \\
            &\ddots &\\
            &  & b^{[n]}_{3,3}
        \end{bmatrix}_{4\ell\times4\ell}
    \end{bmatrix}.\label{common2}
\end{align}
Note that,
\begin{align}
     R_n^{[1]}\begin{bmatrix}
            b^{[n]}_{1,1} &   & \\
            &\ddots &\\
            &  & b^{[n]}_{1,1}        \end{bmatrix}_{4\ell\times4\ell}&=\left(\begin{bmatrix}
        0_{\ell\times\ell} & 0_{\ell\times\ell} & 0_{\ell\times\ell} & \Gamma_n\\
        \Gamma_n & 0_{\ell\times\ell} & 0_{\ell\times\ell} & 0_{\ell\times\ell} \\
        0_{\ell\times\ell} & 0_{\ell\times\ell} & \Gamma_n & 0_{\ell\times\ell}\\
        0_{\ell\times\ell} & \Gamma_n & 0_{\ell\times\ell}&  0_{\ell\times\ell} 
    \end{bmatrix}+\Tilde{Z}^{[1]}\right)\nonumber\\
    &\qquad\times\begin{bmatrix}
        \Gamma_n^{-1}\hat{Z}_{1,1} & 0_{\ell\times\ell} & 0_{\ell\times\ell} & 0_{\ell\times\ell}\\
        0_{\ell\times\ell} & \Gamma_n^{-1}\hat{Z}_{1,1} & 0_{\ell\times\ell} & 0_{\ell\times\ell}\\
        0_{\ell\times\ell} & 0_{\ell\times\ell} & \Gamma_n^{-1}\hat{Z}_{1,1} &  0_{\ell\times\ell}\\
        0_{\ell\times\ell} & 0_{\ell\times\ell} & 0_{\ell\times\ell} & \Gamma_n^{-1}\hat{Z}_{1,1}\\
    \end{bmatrix}\\
    &=\begin{bmatrix}
        0_{\ell\times\ell} & 0_{\ell\times\ell} & 0_{\ell\times\ell} & \hat{Z}_{1,1}\\        
        \hat{Z}_{1,1} & 0_{\ell\times\ell} & 0_{\ell\times\ell} & 0_{\ell\times\ell} \\     0_{\ell\times\ell} & 0_{\ell\times\ell} & \hat{Z}_{1,1} & 0_{\ell\times\ell} \\        
        0_{\ell\times\ell} & \hat{Z}_{1,1} & 0_{\ell\times\ell} & 0_{\ell\times\ell} \\
    \end{bmatrix}+P_{\alpha_n}(1)=P_{\alpha_n}(1)
\end{align}
where $\hat{Z}_{1,1}$ is the submatrix of $\hat{Z}$ in \eqref{eg_blk} consisting of the first $\ell$ rows and first $\ell$ columns, $P_{\alpha_n}(1)$ here are matrices of size $4\ell\times4\ell$, consisting of polynomials in $ \alpha_n$ of degree $1$ and $0_{\ell\times\ell}$ is the all zeros matrix of size $\ell\times\ell$. Next, consider the calculation of,
\begin{align}
    R_n^{[1]}\begin{bmatrix}
            b^{[n]}_{1,3} &   & \\
            &\ddots &\\
            &  & b^{[n]}_{1,3}        \end{bmatrix}_{4\ell\times4\ell}\!\!\!&=\left(\begin{bmatrix}
        0_{\ell\times\ell} & 0_{\ell\times\ell} & 0_{\ell\times\ell} & \Gamma_n\\
        \Gamma_n & 0_{\ell\times\ell} & 0_{\ell\times\ell} & 0_{\ell\times\ell} \\
        0_{\ell\times\ell} & 0_{\ell\times\ell} & \Gamma_n & 0_{\ell\times\ell}\\
        0_{\ell\times\ell} & \Gamma_n & 0_{\ell\times\ell}&  0_{\ell\times\ell} 
    \end{bmatrix}+\Tilde{Z}^{[1]}\right)\nonumber\\
    &\times\left(\begin{bmatrix}
        \Phi & 0_{\ell\times\ell} & 0_{\ell\times\ell} & 0_{\ell\times\ell}\\
        0_{\ell\times\ell} & \Phi & 0_{\ell\times\ell} & 0_{\ell\times\ell}\\
        0_{\ell\times\ell} & 0_{\ell\times\ell} & \Phi & 0_{\ell\times\ell}\\
        0_{\ell\times\ell} & 0_{\ell\times\ell} & 0_{\ell\times\ell} & \Phi
    \end{bmatrix}+\begin{bmatrix}
        \Gamma_n^{-1}\hat{Z}_{1,3} & 0_{\ell\times\ell} & 0_{\ell\times\ell} & 0_{\ell\times\ell}\\
        0_{\ell\times\ell} & \Gamma_n^{-1}\hat{Z}_{1,3} & 0_{\ell\times\ell} & 0_{\ell\times\ell}\\
        0_{\ell\times\ell} & 0_{\ell\times\ell} & \Gamma_n^{-1}\hat{Z}_{1,3} &  0_{\ell\times\ell}\\
        0_{\ell\times\ell} & 0_{\ell\times\ell} & 0_{\ell\times\ell} & \Gamma_n^{-1}\hat{Z}_{1,3}\\
    \end{bmatrix}\right)\\
    &=\begin{bmatrix}
        0_{\ell\times\ell} & 0_{\ell\times\ell} & 0_{\ell\times\ell} & \Gamma_n\\
        \Gamma_n & 0_{\ell\times\ell} & 0_{\ell\times\ell} & 0_{\ell\times\ell} \\
        0_{\ell\times\ell} & 0_{\ell\times\ell} & \Gamma_n & 0_{\ell\times\ell}\\
        0_{\ell\times\ell} & \Gamma_n & 0_{\ell\times\ell} & 0_{\ell\times\ell} \\
    \end{bmatrix}+P_{\alpha_n}(1),
\end{align}
where $\hat{Z}_{1,3}$ is the submatrix of $\hat{Z}$ in \eqref{eg_blk} consisting of the first $\ell$ rows and the column indices given by $2\ell+1$ to $3\ell$ and $P_{\alpha_n}(1)$ is a matrix of size $4\ell\times4\ell$ consisting of polynomials of $\alpha_n$ of degree $1$. Based on similar calculations, we can write the combined permutation reversing matrix $R_n$ in \eqref{common2} as,
\begin{align}
    R_n&=\begin{bmatrix}
        0_{4\ell\times4\ell} & 0_{4\ell\times4\ell} &
        \begin{bmatrix}
        0_{\ell\times\ell} & 0_{\ell\times\ell} & 0_{\ell\times\ell} & \Gamma_n\\
        \Gamma_n & 0_{\ell\times\ell} & 0_{\ell\times\ell} & 0_{\ell\times\ell} \\
        0_{\ell\times\ell} & 0_{\ell\times\ell} & \Gamma_n & 0_{\ell\times\ell}\\
        0_{\ell\times\ell} & \Gamma_n & 0_{\ell\times\ell} & 0_{\ell\times\ell} \\
    \end{bmatrix} \\
    \begin{bmatrix}
        \Gamma_n & 0_{\ell\times\ell} & 0_{\ell\times\ell} & 0_{\ell\times\ell} \\
        0_{\ell\times\ell} & 0_{\ell\times\ell} & \Gamma_n & 0_{\ell\times\ell} \\
        0_{\ell\times\ell} & \Gamma_n & 0_{\ell\times\ell}  & 0_{\ell\times\ell}\\
        0_{\ell\times\ell}  & 0_{\ell\times\ell}&  0_{\ell\times\ell} & \Gamma_n
    \end{bmatrix} & 0_{4\ell\times4\ell} & 0_{4\ell\times4\ell}\\
    0_{4\ell\times4\ell} & \begin{bmatrix}
    0_{\ell\times\ell}  & \Gamma_n & 0_{\ell\times\ell} & 0_{\ell\times\ell} \\
    0_{\ell\times\ell} & 0_{\ell\times\ell} & 0_{\ell\times\ell} & \Gamma_n  \\
    \Gamma_n & 0_{\ell\times\ell} & 0_{\ell\times\ell} & 0_{\ell\times\ell}\\
    0_{\ell\times\ell} & 0_{\ell\times\ell}&  \Gamma_n & 0_{\ell\times\ell} 
    \end{bmatrix} & 0_{4\ell\times4\ell}    
    \end{bmatrix}+P_{\alpha_n}(1),\label{eg3_comb}
\end{align}
where $P_{\alpha_n}(1)$ here is a matrix of size $12\ell\times12\ell$, whose elements are polynomials in $\alpha_n$ of degree 1. Note that the first part of the combined noise added permutation reversing matrix $R_n$ is simply the $1$ in the $i$th row of $\Bar{R}$ (permutation reversing matrix corresponding to the inter-segment permutation $\hat{P}$) replaced by $\Tilde{R}^{[i]}\otimes\Gamma_n$, where $\Tilde{R}^{[i]}$ is the permutation reversing matrix corresponding to the within-segment permutation $\Tilde{P}_i$, for each $i\in\{1,2,3\}$.

In order to download the subpacket corresponding to the permuted pair $(\eta_p,\phi_p)$, each database generates the query given by,
\begin{align}
    Q_n^{[\eta_p,\phi_p]}&=(I_P\otimes\Gamma_n^{-1})\times \sum_{k=1}^\ell R_n(:, (\phi_p-1)\frac{P}{B}\ell+(\eta_p-1)\ell+k),
\end{align}
where $I_P$ is the identity matrix of size $P\times P$. Then, database $n$, $n\in\{1,\dotsc,N\}$ sends the answer corrresponding to the permuted subpacket $(\eta_p,\phi_p)$ as,
\begin{align}
    A_n^{[\eta_p,\phi_p]}=S_n^TQ_n^{[\eta_p,\phi_p]}
    =\frac{1}{f_1-\alpha_n}W_1^{[\eta_r,\phi_r]}+\dots+\frac{1}{f_\ell-\alpha_n}W_\ell^{[\eta_r,\phi_r]}+P_{\alpha_n}(\ell+3)\label{ans_3},
\end{align}
where $P_{\alpha_n}(\ell+3)$ is a polynomial in $\alpha_n$ of degree $\ell+3$. Then, the user can obtain the values of the corresponding real subpacket indicated by $(\eta_r,\phi_r)$, i.e., $W_1^{[\eta_r,\phi_r]},\dotsc,W_{\ell}^{[\eta_r,\phi_r]}$ using all answers received by the $N$ databases as,
\begin{align}
    \begin{bmatrix}
        A_1^{[\eta_p,\phi_p]}\\\vdots\\A_N^{[\eta_p,\phi_p]}
    \end{bmatrix}=\begin{bmatrix}
        \frac{1}{f_1-\alpha_1} & \dotsc & \frac{1}{f_{\ell}-\alpha_1} & 1 & \alpha_1 & \dotsc & \alpha_1^{\ell+3}\\
        \vdots & \vdots & \vdots & \vdots & \vdots & \vdots & \vdots\\ 
        \frac{1}{f_1-\alpha_N} & \dotsc & \frac{1}{f_{\ell}-\alpha_N} & 1 & \alpha_N & \dotsc &\alpha_N^{\ell+3}\\
    \end{bmatrix}
    \begin{bmatrix}        W_1^{[\eta_r,\phi_r]}\\\vdots\\W_{\ell}^{[\eta_r,\phi_r]}\\R_{0:\ell+3}
    \end{bmatrix},\label{mat3}
\end{align}
where each $R_i$ corresponds to the $i$th coefficient of the polynomial $P_{\alpha_n}(\ell+3)$ in \eqref{ans_3}. The equation in \eqref{mat3} is solvable if $N=2\ell+4$, which determines the subpacketization as $\ell=\frac{N-4}{2}$.

As an example, consider the download of the permuted subpacket indicated by $(\eta_p,\phi_p)=(1,3)$, from the same example setting in Fig.~\ref{init_c3_eg}. Database $n$, $n\in\{1,\dotsc,N\}$ creates the query given by,
\begin{align}
    Q_n^{[1,3]}&=(I_{12}\otimes\Gamma_n^{-1})\times \sum_{k=1}^\ell R_n(:, 8\ell+k)\\
    &=\begin{bmatrix}
        \Gamma_n^{-1} & &\\
        & \ddots &\\
        & & \Gamma_n^{-1}
    \end{bmatrix}_{12\ell\times 12\ell}\times\left(\begin{bmatrix}
        0_{\ell}\\ \frac{1}{f_1-\alpha_n}\\\vdots\\\frac{1}{f_\ell-\alpha_n}\\ 0_{10\ell} 
    \end{bmatrix}+\dot{P}_{\alpha_n}(1)\right)=\begin{bmatrix}
        0_{\ell}\\ 1_{\ell}\\ 0_{10\ell} 
    \end{bmatrix}+\begin{bmatrix}
        \begin{bmatrix}
            (f_1-\alpha_n)P_{\alpha_n}(1)\\
            \vdots\\
            (f_\ell-\alpha_n)P_{\alpha_n}(1)
        \end{bmatrix}\\
        \vdots\\
        \begin{bmatrix}
            (f_1-\alpha_n)P_{\alpha_n}(1)\\
            \vdots\\
            (f_\ell-\alpha_n)P_{\alpha_n}(1)
        \end{bmatrix}
    \end{bmatrix}_{L\times1},
\end{align}
where $\dot{P}_{\alpha_n}(1)$ is a vector of size $12\ell\times1$, consisting of polynomials in $\alpha_n$ of degree $1$, and $P_{\alpha_n}(1)$ are polynomials of $\alpha_n$ of degree 1. Note that $0_k$ and $1_k$ refer to all zeros and all ones vectors of size $k\times1$, respectively. The answer corresponding to the above query, sent to the users by database $n$, $n\in\{1,\dots,N\}$ is given by,
\begin{align}
    A_n^{[1,3]}&=S_n^TQ_n^{[1,3]}=\begin{bmatrix}
        \begin{bmatrix}
    \frac{1}{f_{1}-\alpha_n}W_{1}^{[1]}+\sum_{j=0}^{\ell+1}\alpha_n^j Z^{[1]}_{1,j}\\
    \vdots\\
    \frac{1}{f_{\ell}-\alpha_n}W_{\ell}^{[1]}+\sum_{j=0}^{\ell+1}\alpha_n^j Z^{[1]}_{\ell,j}
    \end{bmatrix}\\\vdots\\\begin{bmatrix}
    \frac{1}{f_{1}-\alpha_n}W_{1}^{[12]}+\sum_{j=0}^{\ell+1}\alpha_n^j Z^{[12]}_{1,j}\\
    \vdots\\
    \frac{1}{f_{\ell}-\alpha_n}W_{\ell}^{[12]}+\sum_{j=0}^{\ell+1}\alpha_n^j Z^{[12]}_{\ell,j}
    \end{bmatrix}
    \end{bmatrix}^T\times \left(\begin{bmatrix}
        0_{\ell}\\ 1_{\ell}\\ 0_{10\ell} 
    \end{bmatrix}+\begin{bmatrix}
        \begin{bmatrix}
            (f_1-\alpha_n)P_{\alpha_n}(1)\\
            \vdots\\
            (f_\ell-\alpha_n)P_{\alpha_n}(1)
        \end{bmatrix}\\
        \vdots\\
        \begin{bmatrix}
            (f_1-\alpha_n)P_{\alpha_n}(1)\\
            \vdots\\
            (f_\ell-\alpha_n)P_{\alpha_n}(1)
        \end{bmatrix}
    \end{bmatrix}_{L\times1}\right)\\
    &=\frac{1}{f_1-\alpha_n}W_1^{[2]}+\dots+\frac{1}{f_\ell-\alpha_n}W_\ell^{[2]}+P_{\alpha_n}(\ell+3),
\end{align}
from which the (real) second subpacket of segment 1, i.e., $(\eta_r,\phi_r)=(2,1)$ can be obtained using all answers of the $N$ databases if $N=2\ell+4$ is satisfied, which determines the subpacketization as $\ell=\frac{N-4}{2}$. The resulting reading cost is given by,
\begin{align}
    C_R=\frac{Pr'(N+\log_q B+\log_q\frac{P}{B})}{L}=\frac{Pr'(N+\log_q P)}{P\frac{N-4}{2}}=\frac{2r'(1+\frac{\log_q P}{N})}{1-\frac{4}{N}}.
\end{align}

\emph{Writing Phase:} In the writing phase, the user sends the combined updates, permuted subpacket indices and permuted segment indices of the $Pr$ subpackets with the most significant updates to all databases. Let $(\eta_r^{[i]},\phi_r^{[i]})$, $i\in\{1,\dotsc,Pr\}$ be the real (subpacket, segment) pair corresponding to the $i$th selected subpacket. For each of the $Pr$ selected subpackets, the user generates a combined update bit given by,
\begin{align}\label{comb_3}    U_n^{[\eta_r^{[i]},\phi_r^{[i]}]}=\sum_{k=1}^\ell\prod_{j=1,j\neq k}^\ell (f_j-\alpha_n)\Tilde{\Delta}_{k}^{[\eta_r^{[i]},\phi_r^{[i]}]}+\prod_{j=1}^\ell(f_j-\alpha_n)Z^{[\eta_r^{[i]},\phi_r^{[i]}]},\quad i\in\{1,\dotsc,Pr\},
\end{align}
where $\Tilde{\Delta}_{k}^{[\eta_r^{[i]},\phi_r^{[i]}]}=\frac{\Delta_{k}^{[\eta_r^{[i]},\phi_r^{[i]}]}}{\prod_{j=1,j\neq k}^\ell (f_j-f_k)}$, with $\Delta_{k}^{[\eta_r^{[i]},\phi_r^{[i]}]}$ being the update of the $k$th parameter of subpacket $\eta_r^{[i]}$ of segment $\phi_r^{[i]}$ and $Z^{[\eta_r^{[i]},\phi_r^{[i]}]}$ is a random noise symbol. The user sends the permuted (update, subpacket, segment) tuple given by $(U_n^{[\eta_r^{[i]},\phi_r^{[i]}]},\eta_p^{[i]},\phi_p^{[i]})$ for the $i$th selected subpacket where $U_n^{[\eta_r^{[i]},\phi_r^{[i]}]}$ is the combined update of the subpacket of the form \eqref{comb_3}, $\eta_p^{[i]}$ is the permuted subpacket index obtained by,
\begin{align}
    \eta_p^{[i]}=\tilde{P}^{-1}_{\phi_r^{[i]}}(\eta_r^{[i]})
\end{align}
and $\phi_p^{[i]}$ is the permuted segment index obtained by,
\begin{align}
    \phi_p^{[i]}=\hat{P}^{-1}(\phi_r^{[i]})    
\end{align}
where $\tilde{P}_{\phi_r^{[i]}}$ and $\hat{P}$ are considered to be the one-to-one mappings from $j$ to $\tilde{P}_{\phi_r^{[i]}}(j)$ for $j=1,\dotsc,\frac{P}{B}$ and $i$ to $\hat{P}(i)$, for $i=1,\dotsc,B$, respectively. For the example in Fig.~\ref{init_c3_eg}, assume that a given user wants to update the $Pr$ sparse subpackets identified by the real (subpacket, segment) pairs given by, $(\eta_r,\phi_r)=\{(2,1),(2,2),(3,3)\}$. Based on the within segment permutations given by $\Tilde{P}_1=(2,4,3,1)$, $\Tilde{P}_2=(1,3,2,4)$, $\Tilde{P}_3=(3,1,4,2)$, and the segment-wise permutation given by $\hat{P}=(2,3,1)$, the user sends the following (permuted) information to database $n$, $n\in\{1,\dots,N\}$,
\begin{align}
    (U_n^{[\eta_r,\phi_r]},\eta_p,\phi_p)&=\{(U_n^{[2,1]},\Tilde{P}^{-1}_1(2),\hat{P}^{-1}(1)),(U_n^{[2,2]},\Tilde{P}^{-1}_2(2),\hat{P}^{-1}(2)),(U_n^{[3,3]},\Tilde{P}^{-1}_3(3),\hat{P}^{-1}(3))\}\\
    &=\{(U_n^{[2,1]},1,3),(U_n^{[2,2]},3,1),(U_n^{[3,3]},1,2)\},\label{rec_c3}
\end{align}
where the three $U_n^{[\eta_r,\phi_r]}$ terms are generated as in \eqref{comb_3}. As an illustration, the real (subpacket, segment) pair given by $(2,1)$, is converted to the permuted pair $(1,3)$ as follows. Note that in the segment-wise permutation $\hat{P}=(2,3,1)$, the real segment index $1$ lies in the third position. Therefore, $\phi_r=1$ is converted to $\phi_p=3$. Next, in the permutation corresponding to segment 1 ($\Tilde{P}_1=(2,4,3,1)$), the subpacket index 2 lies in the first position. Therefore, $\eta_r=2$ is converted to $\eta_p=1$. 

Once the databases receive all permuted (update, subpacket, segment) tuples, they construct the permuted update vector as,
\begin{align}\label{perm_upd_vec}
    \tilde{U}_n&=\sum_{i=1}^{Pr} U^{[\eta_r^{[i]},\phi_r^{[i]}]}_n e_P((\phi_p^{[i]}-1)\frac{P}{B}+\eta_p^{[i]}),
\end{align}
where $e_p((\phi_p^{[i]}-1)\frac{P}{B}+\eta_p^{[i]})$ is the all zeros vector of size $P$ with a $1$ at the $(\phi_p^{[i]}-1)\frac{P}{B}+\eta_p^{[i]}$th position. This vector is then scaled by an all ones vector of size $\ell\times1$ (i.e., $1_\ell$) to aid the rest of the calculations. The scaled permuted update vector is given by,
\begin{align}
    \hat{U}_n=\Tilde{U}_n\otimes1_{\ell}=\begin{bmatrix}
        \tilde{U}_n(1)\cdot1_\ell\\
        \vdots\\
        \tilde{U}_n(P)\cdot1_\ell
    \end{bmatrix}.
\end{align}
Then, database $n$, $n\in\{1,\dotsc,N\}$ calculates the incremental update using the combined noise added permutation reversing matrix in \eqref{common} as,
\begin{align}
    \bar{U}_n=R_n\times\hat{U}_n,
\end{align}
which is of the same form as the storage in \eqref{storage_c3_eg}. Therefore, the storage at time $t$, $S_n^{[t]}$ can be updated as,
\begin{align}
    S_n^{[t]}=S_n^{[t-1]}+\bar{U}_n.
\end{align}

For the example considered, the scaled permuted update vector based on the information received by the databases in \eqref{rec_c3} is given by,
\begin{align}
    \hat{U}_n=(U_n^{[2,1]}e_{12}(9)+U_n^{[2,2]}e_{12}(3)+U_n^{[3,3]}e_{12}(5))\otimes1_{\ell}=\begin{bmatrix}
        \begin{bmatrix}            0\cdot1_{\ell}\\0\cdot1_{\ell}\\U_n^{[2,2]}\cdot1_{\ell}\\0\cdot1_{\ell}
        \end{bmatrix}\\
        \begin{bmatrix}            U_n^{[3,3]}\cdot1_{\ell}\\0\cdot1_{\ell}\\0\cdot1_{\ell}\\0\cdot1_{\ell}
        \end{bmatrix}\\
        \begin{bmatrix}            U_n^{[2,1]}\cdot1_{\ell}\\0\cdot1_{\ell}\\0\cdot1_{\ell}\\0\cdot1_{\ell}
        \end{bmatrix}
    \end{bmatrix}
\end{align}
Then, database $n$, $n\in\{1,\dots,N\}$ computes the incremental update using the combined noise added permutation reversing matrix in \eqref{eg3_comb} as,
\begin{align}
    \bar{U}_n&=R_n\times\hat{U}_n\\
    &=\!\!\!\begin{bmatrix}\!
        0_{4\ell\times4\ell}\!\!\!\!\!\! &\!\!\!\!\!\! 0_{4\ell\times4\ell}\!\!\!\!\!\! &\!\!\!\!\!\!
        \begin{bmatrix}
        0_{\ell\times\ell} & 0_{\ell\times\ell} & 0_{\ell\times\ell} & \Gamma_n\\
        \Gamma_n & 0_{\ell\times\ell} & 0_{\ell\times\ell} & 0_{\ell\times\ell} \\
        0_{\ell\times\ell} & 0_{\ell\times\ell} & \Gamma_n & 0_{\ell\times\ell}\\
        0_{\ell\times\ell} & \Gamma_n & 0_{\ell\times\ell} & 0_{\ell\times\ell} \\
    \end{bmatrix}\! \\
    \!\begin{bmatrix}
        \Gamma_n & 0_{\ell\times\ell} & 0_{\ell\times\ell} & 0_{\ell\times\ell} \\
        0_{\ell\times\ell} & 0_{\ell\times\ell} & \Gamma_n &0_{\ell\times\ell} \\
        0_{\ell\times\ell} & \Gamma_n & 0_{\ell\times\ell}  & 0_{\ell\times\ell}\\
        0_{\ell\times\ell}  & 0_{\ell\times\ell}&  0_{\ell\times\ell} & \Gamma_n
    \end{bmatrix}\!\!\!\!\!\! &\!\!\!\!\!\! \!\!\!0_{4\ell\times4\ell}\!\!\!\!\!\! &\!\!\!\!\!\!0_{4\ell\times4\ell}\!\\
    0_{4\ell\times4\ell}\!\!\!\!\!\! &\!\!\!\!\!\! \begin{bmatrix}
        0_{\ell\times\ell}  & \Gamma_n & 0_{\ell\times\ell} & 0_{\ell\times\ell} \\
        0_{\ell\times\ell} & 0_{\ell\times\ell} & 0_{\ell\times\ell} & \Gamma_n  \\
        \Gamma_n & 0_{\ell\times\ell} & 0_{\ell\times\ell} & 0_{\ell\times\ell}\\
        0_{\ell\times\ell} &  0_{\ell\times\ell}&  \Gamma_n &0_{\ell\times\ell} 
    \end{bmatrix}\!\!\!\!\!\! &\!\!\!\!\!\! 0_{4\ell\times4\ell} \!   
    \end{bmatrix}\begin{bmatrix}
        \begin{bmatrix}            0\cdot1_{\ell}\\0\cdot1_{\ell}\\U_n^{[2,2]}\cdot1_{\ell}\\0\cdot1_{\ell}
        \end{bmatrix}\\
        \begin{bmatrix}            U_n^{[3,3]}\cdot1_{\ell}\\0\cdot1_{\ell}\\0\cdot1_{\ell}\\0\cdot1_{\ell}
        \end{bmatrix}\\
        \begin{bmatrix}            U_n^{[2,1]}\cdot1_{\ell}\\0\cdot1_{\ell}\\0\cdot1_{\ell}\\0\cdot1_{\ell}
        \end{bmatrix}
    \end{bmatrix}\nonumber\\
    &\qquad \qquad \qquad +P_{\alpha_n}(\ell+1)\\
    &=\left[0_\ell,\frac{U_n^{[2,1]}}{f_1-\alpha_n},\dots,\frac{U_n^{[2,1]}}{f_\ell-\alpha_n},0_{3\ell},\frac{U_n^{[2,2]}}{f_1-\alpha_n},\dots,\frac{U_n^{[2,2]}}{f_\ell-\alpha_n},0_{4\ell},\frac{U_n^{[3,3]}}{f_1-\alpha_n},\dots,\frac{U_n^{[3,3]}}{f_\ell-\alpha_n},0_\ell,\right]^T\nonumber\\
    &\qquad+P_{\alpha_n}(\ell+1)\\
    &=\left[0_\ell,\frac{\Delta_{1}^{[2,1]}}{f_1-\alpha_n},\dots,\frac{\Delta_{\ell}^{[2,1]}}{f_\ell-\alpha_n},0_{3\ell},\frac{\Delta_{1}^{[2,2]}}{f_1-\alpha_n},\dots,\frac{\Delta_{\ell}^{[2,2]}}{f_\ell-\alpha_n},0_{4\ell},\frac{\Delta_{1}^{[3,3]}}{f_1-\alpha_n},\dots,\frac{\Delta_{\ell}^{[3,3]}}{f_\ell-\alpha_n},0_\ell,\right]^T\nonumber\\
    &\qquad+P_{\alpha_n}(\ell+1)\label{last},
\end{align}
where $P_{\alpha_n}(\ell+1)$ is a vector of size $L\times1$ whose elements are polynomials in $\alpha_n$ of degree $\ell+1$, and \eqref{last} follows from Lemma 1 in \cite{pruw_jpurnal}. Note from \eqref{last} that the (real) subpacket 2 of segment 1, subpacket 2 of segment 2 and subpacket 3 of segment 3 are correctly updated, without revealing these real indices to any of the databases.\footnote{Note that the vector $P_{\alpha_n}(\ell+1)$ in \eqref{last} hides these non-zero updates from the databases.} Since $\bar{U}_n$ is in the same form as the storage in \eqref{storage_c3_eg}, the storage at time $t$ can be updated by adding $\bar{U}_n$ to the existing storage. The resulting writing cost is given by,
\begin{align}
    C_W=\frac{PrN(1+\log_q B+\log_q \frac{P}{B})}{L}=\frac{PrN(1+\log_q P)}{P\frac{N-4}{2}}=\frac{2r(1+\log_q P)}{1-\frac{4}{N}}.
\end{align}
The storage complexities of data, within-segment noise added permutation reversing matrices and the inter-segment noise added permutation reversing matrix are given by $O(L)$, $O(\frac{L^2}{B})$ and $O(\ell^2B^2)=O(N^2B^2)$, respectively. Therefore the storage complexity is $\max\{O(\frac{L^2}{B}),O(N^2B^2)\}$. The information leakage (on the indices of sparse updates) is derived in Section~\ref{info_leak}.

\textbf{Case 4:} In this case, we consider coded storage and smaller permutation reversing matrices to reduce the storage cost. Both within-segment and inter-segment permutations are considered in this case to reduce the information leakage.

\emph{Initialization:} The storage of a single subpacket $s$ in database $n$, $n\in\{1,\dotsc,N\}$ is given by,
\begin{align}
    S_n^{[s]}=\sum_{i=1}^\ell \frac{1}{\alpha_n^i}W_i^{[s]}+\sum_{i=0}^{2\ell} \alpha_n^i Z_{i}^{[s]},
\end{align}
with the same notation used in case 2. Therefore, the storage of a given segment $j$, $j\in\{1,\dotsc,B\}$ is given by,
 \begin{align}\label{storage_c4_eg}
 S_{n,j}=\begin{bmatrix}
    \sum_{i=1}^\ell \frac{1}{\alpha_n^i}W_i^{[1,j]}+\sum_{i=0}^{2\ell} \alpha_n^i Z_{i}^{[1,j]}\\
    \vdots\\
    \sum_{i=1}^\ell \frac{1}{\alpha_n^i}W_i^{[\frac{P}{B},j]}+\sum_{i=0}^{2\ell} \alpha_n^i Z_{i}^{[\frac{P}{B},j]}
    \end{bmatrix},
\end{align}
where $W_i^{[k,j]}$ is the $i$th parameter of subpacket $k$ of segment $j$ and $Z_{i}^{[k,j]}$ are random noise symbols. The segments are stacked one after the other in the order of segment $1$ through segment $B$. As described in case 3, the coordinator sends the within-segment and inter-segment permutations $\Tilde{P}_1$, $\dotsc$, $\Tilde{P}_B$ and $\hat{P}$ to the users and the corresponding noise added permutation reversing matrices given by $R_n^{[1]}$, $\dotsc$, $R_n^{[B]}$ and $\hat{R}_n$ to database $n$, $n\in\{1,\dots,N\}$. The noise added permutation reversing matrices for the within segment permutations $\tilde{P}_i$ are of the form \eqref{reverse2}, and the noise added permutation reversing matrix corresponding to the inter-segment permutation $\hat{P}$ is of the form,
\begin{align}\label{inter4}
    \hat{R}_n=\hat{R}+\alpha_n^\ell Z,
\end{align}
where $\hat{R}$ is the permutation reversing matrix corresponding to $\hat{P}$ and $Z$ is a random noise matrix, both of size $B\times B$. For the example considered in Figure~\ref{init_c3_eg}, the noise added permutation reversing matrices corresponding to $\tilde{P}_1=\{2,4,3,1\}$, $\tilde{P}_2=\{1,3,2,4\}$, $\tilde{P}_3=\{3,1,4,2\}$ and $\hat{P}=\{2,3,1\}$, stored in database $n$, $n\in\{1,\dotsc,N\}$ are given by,
\begin{align}
    R_n^{[1]}=\begin{bmatrix}
        0 & 0 & 0 & 1\\
        1 & 0 & 0 & 0\\
        0 & 0 & 1 & 0\\
        0 & 1 & 0 & 0
    \end{bmatrix}+\alpha_n^{\ell}\bar{Z}^{[1]},\quad R_n^{[2]}=\begin{bmatrix}
        1 & 0 & 0 & 0\\
        0 & 0 & 1 & 0\\
        0 & 1 & 0 & 0\\
        0 & 0 & 0 & 1
    \end{bmatrix}+\alpha_n^{\ell}\bar{Z}^{[2]}, \quad R_n^{[3]}=\begin{bmatrix}
        0 & 1 & 0 & 0\\
        0 & 0 & 0 & 1\\
        1 & 0 & 0 & 0\\
        0 & 0 & 1 & 0
    \end{bmatrix}+\alpha_n^{\ell}\bar{Z}^{[3]}
\end{align}
and \begin{align}\label{Rhat4}
    \hat{R}_n=\begin{bmatrix}
        0 & 0 & 1\\
        1 & 0 & 0\\
        0 & 1 & 0
    \end{bmatrix}+\alpha_n^{\ell} Z.
\end{align}
The initialization stage ends and the coordinator leaves the system once the storage, permutations and the noise added permutation reversing matrices are initialized. 

To aid the calculations in the reading and writing phases described next, the databases compute a combined noise added permutation reversing matrix as described in case 3. This matrix for database $n$, $n\in\{1,\dotsc,N\}$ is given by,
\begin{align}
    R_n&=\begin{bmatrix}
        R_n^{[1]} &  & \\
        & \ddots &  \\
        &  & R_n^{[B]}
    \end{bmatrix}\times (\hat{R}_n\otimes I_{\frac{P}{B}})=\begin{bmatrix}
        R_n^{[1]} &  & \\
        & \ddots &  \\
        &  & R_n^{[B]}
    \end{bmatrix}\times \begin{bmatrix}
        \hat{R}_n(1,1)I_{\frac{P}{B}} & \dotsc & \hat{R}_n(1,B)I_{\frac{P}{B}}\\
        \vdots & \vdots & \vdots\\
        \hat{R}_n(B,1)I_{\frac{P}{B}} & \dotsc & \hat{R}_n(B,B)I_{\frac{P}{B}}\\
    \end{bmatrix}
\end{align}
where $I_{\frac{P}{B}}$ is the identity matrix of size $\frac{P}{B}\times\frac{P}{B}$. The combined matrix $R_n$ places $R_n^{[i]}$ at the position of the $1$ in the $i$th row of $\hat{R}$ in \eqref{inter4}, with added noise. The combined noise added permutation reversing matrix for the example considered in Fig.~\ref{init_c3_eg} for database $n$, $n\in\{1,\dotsc,N\}$ is given by,
\begin{align}
    R_n&=\begin{bmatrix}
        R_n^{[1]} & 0_{4\times4} & 0_{4\times4}\\
        0_{4\times4} & R_n^{[2]} & 0_{4\times4}\\
        0_{4\times4} & 0_{4\times4} & R_n^{[3]}
    \end{bmatrix}\times \begin{bmatrix}
        \hat{R}_n(1,1)I_{4} & \hat{R}_n(1,2)I_{4} & \hat{R}_n(1,3)I_{4}\\
        \hat{R}_n(2,1)I_{4} & \hat{R}_n(2,2)I_{4} & \hat{R}_n(2,3)I_{4}\\
        \hat{R}_n(3,1)I_{4} & \hat{R}_n(3,2)I_{4} & \hat{R}_n(3,3)I_{4}\\
    \end{bmatrix}\\
    &=\begin{bmatrix}
        \begin{bmatrix}
        0 & 0 & 0 & 1\\
        1 & 0 & 0 & 0\\
        0 & 0 & 1 & 0\\
        0 & 1 & 0 & 0
    \end{bmatrix}+\alpha_n^{\ell}\bar{Z}^{[1]} & 0_{4\times4} & 0_{4\times4}\\
    0_{4\times4} & \begin{bmatrix}
        1 & 0 & 0 & 0\\
        0 & 0 & 1 & 0\\
        0 & 1 & 0 & 0\\
        0 & 0 & 0 & 1
    \end{bmatrix}+\alpha_n^{\ell}\bar{Z}^{[2]} & 0_{4\times4}\\
    0_{4\times4} & 0_{4\times4} & \begin{bmatrix}
        0 & 1 & 0 & 0\\
        0 & 0 & 0 & 1\\
        1 & 0 & 0 & 0\\
        0 & 0 & 1 & 0    \end{bmatrix}+\alpha_n^{\ell}\bar{Z}^{[3]}\end{bmatrix}\nonumber\\
    &\quad\times \left(\begin{bmatrix}
        0_{4\times4} & 0_{4\times4} & \begin{bmatrix}
            1 & &\\
            & \ddots &\\
            & & 1
        \end{bmatrix}_{4\times4}\\
        \begin{bmatrix}
            1 & &\\
            & \ddots &\\
            & & 1
        \end{bmatrix}_{4\times4} & 0_{4\times4} & 0_{4\times4}\\
        0_{4\times4} & \begin{bmatrix}
            1 & &\\
            & \ddots &\\
            & & 1
        \end{bmatrix}_{4\times4} & 0_{4\times4}
    \end{bmatrix}+\alpha_n^{\ell}\Tilde{Z}\right)\label{ztilde}\\
    &=\begin{bmatrix}
        0_{4\times4} & 0_{4\times4} & \begin{bmatrix}
        0 & 0 & 0 & 1\\
        1 & 0 & 0 & 0\\
        0 & 0 & 1 & 0\\
        0 & 1 & 0 & 0
    \end{bmatrix}\\
    \begin{bmatrix}
        1 & 0 & 0 & 0\\
        0 & 0 & 1 & 0\\
        0 & 1 & 0 & 0\\
        0 & 0 & 0 & 1
    \end{bmatrix} & 0_{4\times4} & 0_{4\times4}\\
    0_{4\times4} & \begin{bmatrix}
        0 & 1 & 0 & 0\\
        0 & 0 & 0 & 1\\
        1 & 0 & 0 & 0\\
        0 & 0 & 1 & 0    
        \end{bmatrix} & 0_{4\times4}
    \end{bmatrix}+\alpha_n^{\ell}P_{\alpha_n}(\ell),\label{final4}
\end{align}
where $\Tilde{Z}$ in \eqref{ztilde} is a random noise matrix of size $12\times12$, resulted by the noise component of $\hat{R}_n$ and $P_{\alpha_n}(\ell)$ in \eqref{final4} is a matrix of size $12\times12$ with entries consisting of polynomials in $\alpha_n$ of degree $\ell$. Note that the combined noise added permutation reversing matrix in \eqref{final4} places each $R_n^{[i]}$ at the position of the 1 in the $i$th row of the permutation reversing matrix in \eqref{Rhat4} for $i=1,2,3$, with added noise.

\emph{Reading Phase:} As explained in case 3, the databases determine the permuted (subpacket, segment) tuples given by $(\eta_p,\phi_p)$ for the $Pr'$ sparse subpackets and send them to the users. The users obtain the real (subpacket, segment) information of the $(\eta_p,\phi_p)$ tuples from \eqref{decode_seg} and \eqref{decode_sub}. For the example considered in Fig.~\ref{init_c3_eg}, the $(\eta_p,\phi_p)$ and $(\eta_r,\phi_r)$ pairs in \eqref{eg_perm} and \eqref{eg_real} considered in case 3 are valid for case 4 as well.

In order to send the subpacket corresponding to the permuted (subpacket, segment) pair $(\eta_p,\phi_p)$, database $n$, $n\in\{1,\dots,N\}$ creates a query given by,
\begin{align}
    Q_n^{[\eta_p,\phi_p]}&=R_n(:,(\phi_p-1)\frac{P}{B}+\eta_p).
\end{align}
For $(\eta_p,\phi_p)=(1,3)$, the corresponding query is given by,
\begin{align}
    Q_n^{[1,3]}&=R_n(:,9)=\begin{bmatrix}
        \begin{bmatrix}
            0\\1\\0\\0
        \end{bmatrix}\\ 0_8
    \end{bmatrix}+\alpha_n^{\ell}P_{\alpha_n}(\ell),
\end{align}
where $P_{\alpha_n}(\ell)$ here is a vector of size $12\times1$ with entries consisting of polynomials in $\alpha_n$ of degree $\ell$. The databases send the answer to each query corresponding to $(\eta_p,\phi_p)$ as,
\begin{align}\label{ans_4}
    A_n^{[\eta_p,\phi_p]}&=S_n^TQ_n^{[\eta_p,\phi_p]}
    =\sum_{k=1}^\ell \frac{1}{\alpha_n^k}W_k^{[\eta_r,\phi_r]}+P_{\alpha_n}(4\ell),
\end{align}
where $P_{\alpha_n}(4\ell)$ is a polynomial in $\alpha_n$ of degree $4\ell$. The users can obtain the parameters of the corresponding real (subpacket, segment) pair $(\eta_r,\phi_r)$ using,
\begin{align}
    \begin{bmatrix}        A_1^{[\eta_p,\phi_p]}\\\vdots\\A_N^{[\eta_p,\phi_p]}\end{bmatrix}=\begin{bmatrix}
            \frac{1}{\alpha_1^{\ell}} & \dotsc & \frac{1}{\alpha_1} & 1 & \alpha_1 & \dotsc & \alpha_1^{4\ell}\\
            \vdots & \vdots & \vdots & \vdots & \vdots & \vdots & \vdots\\
            \frac{1}{\alpha_N^{\ell}} & \dotsc & \frac{1}{\alpha_N} & 1 & \alpha_N & \dotsc & \alpha_N^{4\ell}
        \end{bmatrix}
        \begin{bmatrix}
            W_{\ell}^{[\eta_r,\phi_r]}\\\vdots\\W_1^{[\eta_r,\phi_r]}\\R_{0:4\ell}
    \end{bmatrix},
\end{align}
where $R_i$ are the coefficients of the polynomial in \eqref{ans_4}, if $N=5\ell+1$ is satisfied. This determines the subpacketization as $\ell=\frac{N-1}{5}$. For the example considered, the answer for $(\eta_p,\phi_p)=(1,3)$ from database $n$, $n\in\{1,\dotsc,N\}$ is given by,
\begin{align}
    A_n^{[1,3]}&=S_n^TQ_n^{[1,3]}=
    \begin{bmatrix}
    \begin{bmatrix}
    \sum_{i=1}^\ell \frac{1}{\alpha_n^i}W_i^{[1,1]}+\sum_{i=0}^{2\ell} \alpha_n^i Z_{i}^{[1,1]}\\
    \vdots\\
    \sum_{i=1}^\ell \frac{1}{\alpha_n^i}W_i^{[4,1]}+\sum_{i=0}^{2\ell} \alpha_n^i Z_{i}^{[4,1]}
    \end{bmatrix}\\
    \begin{bmatrix}
    \sum_{i=1}^\ell \frac{1}{\alpha_n^i}W_i^{[1,2]}+\sum_{i=0}^{2\ell} \alpha_n^i Z_{i}^{[1,2]}\\
    \vdots\\
    \sum_{i=1}^\ell \frac{1}{\alpha_n^i}W_i^{[4,2]}+\sum_{i=0}^{2\ell} \alpha_n^i Z_{i}^{[4,2]}
    \end{bmatrix}\\
    \begin{bmatrix}
    \sum_{i=1}^\ell \frac{1}{\alpha_n^i}W_i^{[1,3]}+\sum_{i=0}^{2\ell} \alpha_n^i Z_{i}^{[1,3]}\\
    \vdots\\
    \sum_{i=1}^\ell \frac{1}{\alpha_n^i}W_i^{[4,3]}+\sum_{i=0}^{2\ell} \alpha_n^i Z_{i}^{[4,3]}
    \end{bmatrix}
    \end{bmatrix}^T \left(\begin{bmatrix}
        \begin{bmatrix}
            0\\1\\0\\0
        \end{bmatrix}\\ 0_8
    \end{bmatrix}+\alpha_n^{\ell}P_{\alpha_n}(\ell)\right)\\
    &=\sum_{i=1}^\ell \frac{1}{\alpha_n^i}W_i^{[2,1]}+P_{\alpha_n}(4\ell).    
\end{align}
The users can obtain the parameters of the (real) second subpacket of segment 1 (since the real indices corresponding to permuted $(\eta_p,\phi_p)=(1,3)$ are $(\eta_r,\phi_r)=(2,1)$ from \eqref{eg_perm} and \eqref{eg_real}.) using the $N$ answers received if $N=5\ell+1$ is satisfied. This defines the subpacketization for case 4 as $\ell=\frac{N-1}{5}$. The resulting reading cost is given by,
\begin{align}
    C_R=\frac{Pr'(N+\log_q B+\log_q\frac{P}{B})}{L}=\frac{Pr'(N+\log_q P)}{P\frac{N-1}{5}}=\frac{5r'(1+\frac{\log_q P}{N})}{1-\frac{1}{N}}.
\end{align}

\emph{Writing phase:} Similar to case 3, the user selects the $Pr$ subpackets with the most significant updates and let $(\eta_r^{[i]},\phi_r^{[i]})$, $i\in\{1,\dotsc,Pr\}$ be the real (subpacket, segment) information of the $i$th selected subpacket. For each such subpacket, the user generates a combined update (single symbol) given by,
\begin{align}\label{comb4}
    U_n^{[\eta_r^{[i]},\phi_r^{[i]}]}=\sum_{k=1}^\ell \frac{1}{\alpha_n^k}\Delta_{k}^{[\eta_r^{[i]},\phi_r^{[i]}]}+Z^{[\eta_r^{[i]},\phi_r^{[i]}]},
\end{align}
where $\Delta_{k}^{[\eta_r^{[i]},\phi_r^{[i]}]}$ is the update of the $k$th parameter of subpacket $\eta_r^{[i]}$ of segment $\phi_r^{[i]}$, and $Z^{[\eta_r^{[i]},\phi_r^{[i]}]}$ is a random noise symbol. The user sends the permuted (update, subpacket, segment) tuple given by $(U_n^{[\eta_r^{[i]},\phi_r^{[i]}]},\eta_p^{[i]},\phi_p^{[i]})$ for the $i$th sparse subpacket for $i\in\{1,\dotsc,Pr\}$ where $U_n^{[\eta_r^{[i]},\phi_r^{[i]}]}$ is the combined update of the subpacket of the form \eqref{comb4}, $\eta_p^{[i]}$ is the permuted subpacket index obtained by $\eta_p^{[i]}=\tilde{P}^{-1}_{\phi_r^{[i]}}(\eta_r^{[i]})$ and $\phi_p^{[i]}$ is the permuted segment index obtained by $\phi_p^{[i]}=\hat{P}^{-1}(\phi_r^{[i]})$, with the same notation used in the description of case 3. For the example considered, assume that a user wants to update real (subpacket, segment) pairs given by $(\eta_r,\phi_r)=\{(2,1),(2,2),(3,3)\}$. Based on the within-segment permutations given by $\Tilde{P}_1=(2,4,3,1)$, $\Tilde{P}_2=(1,3,2,4)$, $\Tilde{P}_3=(3,1,4,2)$, and the inter-segment permutation given by $\hat{P}=(2,3,1)$, the user sends the following (permuted) information to database $n$, $n\in\{1,\dots,N\}$, as described in case 3, 
\begin{align}\label{rec_c4}
    (U_n,\eta_p,\phi_p)=\{(U_n^{[2,1]},1,3),(U_n^{[2,2]},3,1),(U_n^{[3,3]},1,2)\}.
\end{align}
Based on the permuted sparse update tuples $(U_n^{[\eta_r^{[i]},\phi_r^{[i]}]},\eta_p^{[i]},\phi_p^{[i]})$, $i\in\{1,\dotsc,Pr\}$ received, database $n$, $n\in\{1,\dotsc,N\}$ constructs the permuted update vector given in \eqref{perm_upd_vec}. Then, database $n$, $n\in\{1,\dotsc,N\}$ calculates the permutation-reversed incremental update as, 
\begin{align}
    \bar{U}_n=R_n\tilde{U}_n,
\end{align}
which is of the same form as the storage in \eqref{storage_c4_eg}. Therefore, the storage at time $t$ can be updated as,
\begin{align}
    S_n^{[t]}=S_n^{[t-1]}+\bar{U}_n.
\end{align}
For the example considered, the permuted update vector from \eqref{perm_upd_vec} is given by,
\begin{align}
    \tilde{U}_n=[0,0,U_n^{[2,2]},0,U_n^{[3,3]},0,0,0,U_n^{[2,1]},0,0,0]^T.
\end{align}
Then, each database calculates the permutation-reversed incremental update as,
\begin{align}
    \bar{U}_n&=R_n\tilde{U}_n=\left(\begin{bmatrix}
        0_{4\times4} & 0_{4\times4} & \begin{bmatrix}
        0 & 0 & 0 & 1\\
        1 & 0 & 0 & 0\\
        0 & 0 & 1 & 0\\
        0 & 1 & 0 & 0
    \end{bmatrix}\\
    \begin{bmatrix}
        1 & 0 & 0 & 0\\
        0 & 0 & 1 & 0\\
        0 & 1 & 0 & 0\\
        0 & 0 & 0 & 1
    \end{bmatrix} & 0_{4\times4} & 0_{4\times4}\\
    0_{4\times4} & \begin{bmatrix}
        0 & 1 & 0 & 0\\
        0 & 0 & 0 & 1\\
        1 & 0 & 0 & 0\\
        0 & 0 & 1 & 0    
        \end{bmatrix} & 0_{4\times4}
    \end{bmatrix}+\alpha_n^{\ell}P_{\alpha_n}(\ell)\right)\times\begin{bmatrix}        0\\0\\U_n^{[2,2]}\\0\\U_n^{[3,3]}\\0\\0\\0\\U_n^{[2,1]}\\0\\0\\0
    \end{bmatrix}\\    &=[0,U_n^{[2,1]},0,0,0,U_n^{[2,2]},0,0,0,0,U_n^{[3,3]},0]^T+P_{\alpha_n}(2\ell)\\
    &=\left[0,\sum_{i=1}^\ell \frac{1}{\alpha_n^i}\Delta_{i}^{[2,1]},0,0,0,\sum_{i=1}^\ell \frac{1}{\alpha_n^i}\Delta_{i}^{[2,2]},0,0,0,0,\sum_{i=1}^\ell \frac{1}{\alpha_n^i}\Delta_{i}^{[3,3]},0\right]^T+P_{\alpha_n}(2\ell)\label{last_c4}
\end{align}
where $P_{\alpha_n}(2\ell)$ is a vector of size $12\times1$, consisting of polynomials in $\alpha_n$ of degree $2\ell$. Note that the (real) subpacket 2 of segment 1, subpacket 2 of segment 2 and subpacket 3 of segment 3 ($(\eta_r,\phi_r)=\{(2,1),(2,2),(3,3)\}$) are correctly placed in \eqref{last_c4}, without revealing the real indices to the databases.\footnote{The sparse updates in the first part of \eqref{last_c4} are hidden from the databases by the noise vector $P_{\alpha_n}(2\ell)$.} Since the incremental update in \eqref{last_c4} is of the same form as \eqref{storage_c4_eg}, it is directly added to the existing storage to obtain the updated storage. The writing cost is given by,
\begin{align}
    C_W=\frac{PrN(1+\log_q B+\log_q \frac{P}{B})}{L}=\frac{PrN(1+\log_q P)}{P\frac{N-1}{5}}=\frac{5r(1+\log_q P)}{1-\frac{1}{N}}.
\end{align}
The storage complexities of data, noise added within and inter-segment permutation reversing matrix are given by $O(P)=O(\frac{L}{N})$, $O(\frac{P^2}{B})=O(\frac{L^2}{N^2B})$ and $O(B^2)$, respectively. Therefore, the storage complexity is $\max\{O(\frac{L^2}{N^2B}),O(B^2)\}$. The information leakage on the indices of the sparse updates is derived in Section~\ref{info_leak}.

\subsection{Information Leakage}\label{info_leak}

In this section, we quantify the information leakage of the four cases for a given FL setting with $N$ databases, $P$ subpackets, $B$ segments and uplink and downlink sparsficiation rates give by $r$, $r'$. In the proposed scheme, the users send no information to the databases in the reading phase, and the databases determine the sparse set of subpackets and send them to the users. Therefore, there is no information leakage in the reading phase. In the writing phase, the users send $Pr$ permuted tuples of the form (update, subpacket, segment) to databases, from which a given amount of information about the sparse subpacket indices updated by a given user is allowed to leak. 

There are two types of information within a given user's uploads that leak information about the user's local data, namely, 1) values of sparse updates, 2) positions of sparse updates. Note that in the proposed scheme, a random noise symbol is added to all combined sparse updates sent to the databases in all four cases. Therefore, from Shannon's one time pad theorem, the combined update values sent by the user to all databases are random noise symbols which are independent of the values of the sparse updates included in it. Therefore, the amount of information leaked by the values of the sparse updates in this work is zero. 

From the set of permuted (update, subpacket, segment) tuples sent by a given user at time $t$, only the (subpacket, segment) pairs may possibly contain information about the real positions of the sparse updates, since the \emph{update} component is simply random noise that is independent of the real positions of the sparse updates. Let $Y^{[t]}$ be the set of $Pr$ subpacket indices corresponding to the set of permuted (subpacket, segment) pairs sent by a given user at time $t$. Let $X^{[t]}$ be the set of real indices of the $Pr$ sparse subpackets of the model, chosen to be updated by the user at time $t$. Therefore, the amount of information leaked to the databases about the real positions of the sparse updates at time $t$ is quantified by the mututal information between $X^{[t]}$ and $Y^{[t]}$, i.e., $I(X^{[t]};Y^{[t]})$. In this section, we quantify this mutual information for all four cases.

\textbf{Cases 1 and 2:} In cases 1 and 2, permutations exist only within each segment and not among segments. To simplify the notation, we drop the time index in the following calculation, and assume that each $X$ and $Y$ correspond to real and permuted quantities at time $t$. In order to quantify $I(X;Y)$, we first derive the following conditional probability.
\begin{align}    P(X=x&|Y=y)=\frac{\sum_{\Tilde{p}_1,\dots,\Tilde{p}_B}P(X=x,Y=y,\Tilde{P}_1=\Tilde{p}_1,\dots,\Tilde{P}_B=\Tilde{p}_B)}{P(Y=y)}\\
&=\frac{\sum_{\Tilde{p}_1,\dots,\Tilde{p}_B}P(Y=y|X=x,\Tilde{P}_1=\Tilde{p}_1,\dots,\Tilde{P}_B=\Tilde{p}_B)P(X=x)\prod_{i=1}^B P(\Tilde{P}_i=\Tilde{p}_i)}{P(Y=y)}\label{ind}\\
&=\frac{P(X=x)\sum_{\Tilde{p}_1,\dots,\Tilde{p}_B} \mathbf{1}_{\{Y=y,X=x,\Tilde{P}_1=\Tilde{p}_1,\dots,\Tilde{P}_B=\Tilde{p}_B\}}\prod_{i=1}^B P(\Tilde{P}_i=\Tilde{p}_i)}{P(Y=y)}\\
&=\begin{cases}
    \frac{P(X=x)\prod_{i=1}^B \hat{y}_i! \prod_{i=1}^B (\frac{P}{B}-\hat{y}_i)!}{\left(\frac{P}{B}\right)!^B P(Y=y)}, & \text{for $x,y: \hat{x}_i=\hat{y}_i$, $\forall$ $i$ }\\
    0, & \text{otherwise}
\end{cases}\label{count}\\
&=\begin{cases}
    \frac{P(X=x)}{P(Y=y)}\prod_{i=1}^B\frac{\hat{y}_i!(\frac{P}{B}-\hat{y}_i)!}{\left(\frac{P}{B}\right)!}, & \text{for $x,y: \hat{x}_i=\hat{y}_i$, $\forall$ $i$ }\\
    0, & \text{otherwise}
\end{cases}\\
&=\begin{cases}
    \frac{P(X=x)}{P(Y=y)}\prod_{i=1}^B\frac{1}{\binom{P/B}{\hat{y}_i}}, & \text{for $x,y:\hat{x}_i=\hat{y}_i$, $\forall$ $i$ }\\
    0, & \text{otherwise}
\end{cases}\label{step}
\end{align}
where $\hat{x}_i$ and $\hat{y}_i$ are the numbers of real and permuted sparse subpackets in segment $i$, respectively, and \eqref{ind} is obtained by the mutual independence of the permutations of the $B$ segments and the real positions of the sparse updates. \eqref{count} is derived by counting the number of all possible permutations that result in the given $Y=y$ from the given $X=x$. Next, we compute the probability,
\begin{align}    P(Y=y)&=\sum_{\Tilde{p}_1,\dots,\Tilde{p}_B}\sum_{x} P(Y=y,X=x,\Tilde{P}_1=\Tilde{p}_1,\dots,\Tilde{P}_B=\Tilde{p}_B)\\
    &=\sum_{\Tilde{p}_1,\dots,\Tilde{p}_B}\sum_{x} P(Y=y|X=x,\Tilde{P}_1=\Tilde{p}_1,\dots,\Tilde{P}_B=\Tilde{p}_B)P(X=x)\prod_{i=1}^B P(\Tilde{P}_i=\Tilde{p}_i)\\
    &=\sum_{x}P(X=x)\sum_{\Tilde{p}_1,\dots,\Tilde{p}_B} \mathbf{1}_{\{Y=y,X=x,\Tilde{P}_1=\Tilde{p}_1,\dots,\Tilde{P}_B=\Tilde{p}_B\}}\prod_{i=1}^B P(\Tilde{P}_i=\Tilde{p}_i)\\
    &=\sum_{x:\hat{x}_i=\hat{y}_i,\forall i}P(X=x)\frac{\prod_{i=1}^B \hat{y}_i! \prod_{i=1}^B (\frac{P}{B}-\hat{y}_i)!}{\frac{1}{\left(\frac{P}{B}\right)!^B}}\\
    &=\prod_{i=1}^B\frac{1}{\binom{P/B}{\hat{y}_i}}\sum_{x:\hat{x}_i=\hat{y}_i,\forall i} P(X=x)
\end{align}
for each $y$ such that $\sum_{i=1}^B \hat{y}_i=Pr$. Therefore, from \eqref{step},
\begin{align}
    P(X=x&|Y=y)=\begin{cases}
        \frac{P(X=x)}{\sum_{x:\hat{x}_i=\hat{y}_i,\forall i} P(X=x)}, & \text{for $x,y:\hat{x}_i=\hat{y}_i$, $\forall i$}\\
        0, & \text{otherwise}
    \end{cases}
\end{align}.
Then,
\begin{align}
    H(X|Y=y)&=-\sum_{x}P(X=x|Y=y)\log P(X=x|Y=y)\\
    &=-\sum_{x:\hat{x}_i=\hat{y}_i,\forall i} \frac{P(X=x)}{\sum_{x:\hat{x}_i=\hat{y}_i,\forall i} P(X=x)}\log \frac{P(X=x)}{\sum_{x:\hat{x}_i=\hat{y}_i,\forall i} P(X=x)},
\end{align}
from which we obtain,
\begin{align}
    H(X&|Y)=\sum_{y} P(Y=y)H(X|Y=y)\\
    &=-\sum_{y} \left(\prod_{i=1}^B\frac{1}{\binom{P/B}{\hat{y}_i}}\sum_{x:\hat{x}_i=\hat{y}_i,\forall i} P(X=x)\right)\nonumber\\
    &\qquad \times\sum_{x:\hat{x}_i=\hat{y}_i,\forall i} \frac{P(X=x)}{\sum_{x:\hat{x}_i=\hat{y}_i,\forall i} P(X=x)}\log \frac{P(X=x)}{\sum_{x:\hat{x}_i=\hat{y}_i,\forall i} P(X=x)}\\
    &=-\sum_{\hat{y}:\sum_{i=1}^B\hat{y}_i=Pr} \prod_{i=1}^B \binom{P/B}{\hat{y}_i}\left(\prod_{i=1}^B\frac{1}{\binom{P/B}{\hat{y}_i}}\sum_{x:\hat{x}_i=\hat{y}_i,\forall i} P(X=x)\right)\nonumber\\
    &\qquad \times\sum_{x:\hat{x}_i=\hat{y}_i,\forall i} \frac{P(X=x)}{\sum_{x:\hat{x}_i=\hat{y}_i,\forall i} P(X=x)}\log \frac{P(X=x)}{\sum_{x:\hat{x}_i=\hat{y}_i,\forall i} P(X=x)}\\
    &=-\sum_{\hat{y}:\sum_{i=1}^B\hat{y}_i=Pr} \sum_{x:\hat{x}_i=\hat{y}_i,\forall i} P(X=x)\sum_{x:\hat{x}_i=\hat{y}_i,\forall i} \frac{P(X=x)}{\sum_{x:\hat{x}_i=\hat{y}_i,\forall i} P(X=x)}\log \frac{P(X=x)}{\sum_{x:\hat{x}_i=\hat{y}_i,\forall i} P(X=x)}\\
    &=-\sum_{\hat{x}:\sum_{i=1}^B\hat{x}_i=Pr}\sum_{x\in\hat{x}} P(X=x)\sum_{x\in\hat{x}} \frac{P(X=x)}{\sum_{x\in\hat{x}} P(X=x)}\log \frac{P(X=x)}{\sum_{x\in\hat{x}} P(X=x)}\label{step2},
\end{align}
where $\hat{y}=(\hat{y}_1,\dots,\hat{y}_B)$ and $\hat{x}=(\hat{x}_1,\dots,\hat{x}_B)$ are specific realizations of the numbers of permuted and real sparse subpackets in each of the $B$ segments, respectively, and $x\in\hat{X}$ corresponds to each realization of $X$ that result in $\hat{x}_1,\dots,\hat{x}_B$ numbers of sparse subpackets in the $B$ segments. In general, the random variable $\hat{X}=(\hat{X}_1,\dots,\hat{X}_B)$ represents the numbers of (real) sparse subpackets updated by the user in each of the $B$ segments, such that they sum up to $Pr$. Note that we do not assume any specific distribution of $X$ in this calculation. Observing that $\sum_{x\in\hat{x}} P(X=x)=P(\hat{X}=\hat{x})$, the conditional entropy in \eqref{step2} simplifies to,
\begin{align}
    H(X|Y)&=\sum_{\hat{x}:\sum_{i=1}^B\hat{x}_i=Pr} P(\hat{X}=\hat{x})\log P(\hat{X}=\hat{x})-\sum_{\hat{x}:\sum_{i=1}^B\hat{x}_i=Pr}\sum_{x\in\hat{x}} P(X=x)\log P(X=x)\\
    &=-H(\hat{X})+H(X),
\end{align}
since all realizations of $X$ satisfy $\sum_{i=1}^B\hat{x}_i=Pr$, based on the given uplink sparsification rate. Therefore,
\begin{align}
    I(X;Y)=H(X)-H(X|Y)=H(\hat{X})=H(\hat{X}_1,\dots,\hat{X}_B).
\end{align}

\textbf{Cases 3 and 4:} In cases 3 and 4, we consider permutations within segments as well as among segments to reduce the information leakage further. Recall that the permutations within the $B$ segments are denoted by $\{\Tilde{P}_i\}_{i=1}^B$ and the permutation among segments is denoted by $\hat{P}$. Similar to the above calculation, in order to calculate the information leakage $I(X;Y)$, we first compute the conditional distribution given by,
\begin{align}    
P(X=x|Y=y)&=\left(\sum_{\Tilde{p}_1,\dots,\Tilde{p}_B}\sum_{\hat{p}} P(Y=y|X=x,\hat{P}=\hat{p},\Tilde{P}_1=\Tilde{p}_1,\dots,\Tilde{P}_B=\Tilde{p}_B)\right.\nonumber\\
&\qquad \qquad \qquad \left.\times P(X=x)P(\hat{P}=\hat{p})\prod_{i=1}^B P(\Tilde{P}_i=\Tilde{p}_i)\right)/P(Y=y)\\
&=\frac{P(X=x)}{P(Y=y)}\frac{1}{B!}\frac{1}{(P/B)!^B}\sum_{\Tilde{p}_1,\dots,\Tilde{p}_B}\sum_{\hat{p}}\mathbf{1}_{\{Y=y,X=x,\hat{P}=\hat{p},\Tilde{P}_1=\Tilde{p}_1,\dots,\Tilde{P}_B=\Tilde{p}_B\}}\\
&=\begin{cases}
    \frac{P(X=x)}{P(Y=y)}\frac{1}{B!}\frac{1}{(P/B)!^B}\prod_{i=1}^B \hat{y}_i! (P/B-\hat{y}_i)!\prod_{i=1}^B K_{\hat{y}_i}!, & \text{for  $x,y:\{\hat{x}\}=\{\hat{y}\}$}\\
    0, & \text{otherwise}
\end{cases}\\
&=\begin{cases}
    \frac{P(X=x)}{P(Y=y)}\frac{1}{B!}\prod_{i=1}^B \frac{1}{\binom{P/B}{\hat{y}_i}} K_{\hat{y}_i}!, & \text{for  $x,y:\{\hat{x}\}=\{\hat{y}\}$}\\
    0, & \text{otherwise}
\end{cases}\label{cond2}
\end{align}
where $K_{\hat{y}_i}=\sum_{j=1}^B \mathbf{1}_{\hat{y}_j=\hat{y}_i}$, (i.e., number of segments with equal number of sparse subpackets as that of segment $i$) and the notation $\{\hat{x}\}=\{\hat{y}\}$ implies that the two sets $\hat{x}$ and $\hat{y}$ are the same, irrespective of their order, i.e., if $\{\hat{x}\}=\{\hat{y}\}$, for each $\hat{x}_i\in\hat{x}$, there exist some $\hat{y}_j\in\hat{y}$, such that $\hat{x}_i=\hat{y}_j$, and vice versa. Next we calculate $P(Y=y)$ for any $y$ such that $\sum_{i=1}^{Pr} \hat{y}_i=Pr$ as,
\begin{align}    P(Y=y)&=\sum_{x}\sum_{\Tilde{p}_1,\dots,\Tilde{p}_B}\sum_{\hat{p}} P(Y=y|X=x,\hat{P}=\hat{p},\Tilde{P}_1=\Tilde{p}_1,\dots,\Tilde{P}_B=\Tilde{p}_B)\nonumber\\
&\qquad \times P(X=x)P(\hat{P}=\hat{p})\prod_{i=1}^B P(\Tilde{P}_i=\Tilde{p}_i)\\
&=\sum_x P(X=x)\frac{1}{B!}\frac{1}{(P/B)!^B}\sum_{\Tilde{p}_1,\dots,\Tilde{p}_B}\sum_{\hat{p}}\mathbf{1}_{\{Y=y,X=x,\hat{P}=\hat{p},\Tilde{P}_1=\Tilde{p}_1,\dots,\Tilde{P}_B=\Tilde{p}_B\}}\\
&=\frac{1}{B!}\prod_{i=1}^B \frac{1}{\binom{P/B}{\hat{y}_i}} K_{\hat{y}_i}!\sum_{x:\{\hat{x}\}=\{\hat{y}\}} P(X=x)
\end{align}
Therefore, from \eqref{cond2},
\begin{align}
    P(X=x|Y=y)&=\begin{cases}
    \frac{P(X=x)}{\sum_{x:\{\hat{x}\}=\{\hat{y}\}} P(X=x)}, &\text{for  $x,y:\{\hat{x}\}=\{\hat{y}\}$}\\
    0, & \text{otherwise}
\end{cases}.
\end{align}
Then,
\begin{align}
    H(X&|Y=y)=-\sum_{x}P(X=x|Y=y)\log P(X=x|Y=y)\\
    &=-\sum_{x:\{\hat{x}\}=\{\hat{y}\}} \frac{P(X=x)}{\sum_{x:\{\hat{x}\}=\{\hat{y}\}} P(X=x)}\log \frac{P(X=x)}{\sum_{x:\{\hat{x}\}=\{\hat{y}\}} P(X=x)}\\
    &=\log \left(\sum_{x:\{\hat{x}\}=\{\hat{y}\}} P(X=x)\right)-\frac{1}{\sum_{x:\{\hat{x}\}=\{\hat{y}\}} P(X=x)}\sum_{x:\{\hat{x}\}=\{\hat{y}\}} P(X=x)\log P(X=x),
\end{align}
from which we obtain,
\begin{align}
    H(X&|Y)=\sum_{y}P(Y=y)H(X|Y=y)\\
    &=\sum_{y} \frac{1}{B!}\prod_{i=1}^B \frac{1}{\binom{P/B}{\hat{y}_i}} K_{\hat{y}_i}!\sum_{x:\{\hat{x}\}=\{\hat{y}\}} P(X=x)\log \left(\sum_{x:\{\hat{x}\}=\{\hat{y}\}} P(X=x)\right)\nonumber\\
    &\qquad-\sum_{y} \frac{1}{B!}\prod_{i=1}^B \frac{1}{\binom{P/B}{\hat{y}_i}} K_{\hat{y}_i}!\sum_{x:\{\hat{x}\}=\{\hat{y}\}} P(X=x)\frac{1}{\sum_{x:\{\hat{x}\}=\{\hat{y}\}} P(X=x)}\nonumber\\
    &\qquad\qquad \times\sum_{x:\{\hat{x}\}=\{\hat{y}\}} P(X=x)\log P(X=x)\\
    &=\frac{1}{B!}\sum_{\hat{y}:\sum_{i=1}^B\hat{y}_i=Pr} \prod_{i=1}^B \binom{P/B}{\hat{y}_i}\prod_{i=1}^B \frac{1}{\binom{P/B}{\hat{y}_i}} K_{\hat{y}_i}!\sum_{x:\{\hat{x}\}=\{\hat{y}\}} P(X=x)\log \left(\sum_{x:\{\hat{x}\}=\{\hat{y}\}} P(X=x)\right)\nonumber\\
    &\quad -\frac{1}{B!}\sum_{\hat{y}:\sum_{i=1}^B\hat{y}_i=Pr} \prod_{i=1}^B \binom{P/B}{\hat{y}_i}\prod_{i=1}^B \frac{1}{\binom{P/B}{\hat{y}_i}} K_{\hat{y}_i}!\sum_{x:\{\hat{x}\}=\{\hat{y}\}} P(X=x)\log P(X=x)\\
    &=\frac{1}{B!}\sum_{\hat{y}:\sum_{i=1}^B\hat{y}_i=Pr} \prod_{i=1}^B K_{\hat{y}_i}!\sum_{x:\{\hat{x}\}=\{\hat{y}\}} P(X=x)\log \left(\sum_{x:\{\hat{x}\}=\{\hat{y}\}} P(X=x)\right)\nonumber\\
    &\quad -\frac{1}{B!}\sum_{\hat{y}:\sum_{i=1}^B\hat{y}_i=Pr}  \prod_{i=1}^BK_{\hat{y}_i}!\sum_{x:\{\hat{x}\}=\{\hat{y}\}} P(X=x)\log P(X=x)\\
    &=\frac{1}{B!}\sum_{\tilde{y}:\sum_{i=1}^B\hat{y}_i=Pr} \frac{B!}{\prod_{i=1}^BK_{\hat{y}_i}!}\prod_{i=1}^BK_{\hat{y}_i}!\sum_{x:\{\hat{x}\}=\{\hat{y}\}} P(X=x)\log \left(\sum_{x:\{\hat{x}\}=\{\hat{y}\}} P(X=x)\right)\nonumber\\
    &\quad -\frac{1}{B!}\sum_{\tilde{y}:\sum_{i=1}^B\hat{y}_i=Pr} \frac{B!}{\prod_{i=1}^BK_{\hat{y}_i}!} \prod_{i=1}^BK_{\hat{y}_i}!\sum_{x:\{\hat{x}\}=\{\hat{y}\}} P(X=x)\log P(X=x)\label{tild}\\
    &=\sum_{\tilde{y}:\sum_{i=1}^B\hat{y}_i=Pr} \sum_{x:\{\hat{x}\}=\{\hat{y}\}} P(X=x)\log \left(\sum_{x:\{\hat{x}\}=\{\hat{y}\}} P(X=x)\right)\nonumber\\
    &\quad -\sum_{\tilde{y}:\sum_{i=1}^B\hat{y}_i=Pr} \sum_{x:\{\hat{x}\}=\{\hat{y}\}} P(X=x)\log P(X=x),
\end{align}
where $\tilde{y}$ introduced in \eqref{tild} and $\tilde{x}$ are the realizations of corresponding random variables $\Tilde{Y}$ and $\Tilde{X}$ representing all distinct sets of $\hat{y}$ and $\hat{x}$, respectively. For example, if $B=2$, $(1,2)$ and $(2,1)$ are considered to be two different realizations of $\hat{y}$ (or $\hat{x}$), while it is the same realization of $\tilde{y}$ (or $\tilde{x}$). Moreover,
\begin{align}
    P(\tilde{X}=\tilde{x})&=\sum_{x\in\tilde{x}} P(X=x)\\
    &=\sum_{\text{all permutations of } x\in\hat{x}} P(X=x)
\end{align}
With this notation, the above entropy is further simplified to,
\begin{align}
    H(X|Y)&=\sum_{\tilde{y}:\sum_{i=1}^B\hat{y}_i=Pr} \sum_{x:\{\hat{x}\}=\{\hat{y}\}} P(X=x)\log \left(\sum_{x:\{\hat{x}\}=\{\hat{y}\}} P(X=x)\right)\nonumber\\
    &\qquad\qquad -\sum_{\tilde{y}:\sum_{i=1}^B\hat{y}_i=Pr} \sum_{x:\{\hat{x}\}=\{\hat{y}\}} P(X=x)\log P(X=x)\\
    &=\sum_{\tilde{y}:\sum_{i=1}^B\hat{y}_i=Pr} P(\tilde{X}=\tilde{y})\log P(\tilde{X}=\tilde{y})+H(X)\\
    &=-H(\tilde{X})+H(X).
\end{align}
Therefore, the information leakage is given by,
\begin{align}
    I(X;Y)=H(X)-H(X|Y)=H(\tilde{X})=H(\tilde{X}_1,\dots,\tilde{X}_B).
\end{align}

\section{Conclusions}\label{conclusion}
In this work, we considered the problem of private FL with top $r$ sparsification. In FL with top $r$ sparsification, the values and the positions of the sparse updates leak information about the user's private data. We proposed four schemes with different properties to perform FL with top $r$ sparsification without revealing the values or the positions of the sparse updates to the databases. The schemes follow a permutation technique which requires a large storage cost. To this end, we generalized the schemes to incur a reduced storage cost at the expense of a certain amount of information leakage, using a model segmentation mechanism. In general, this work presents the tradeoff between the communication cost, storage complexity and information leakage in private FL with top $r$ sparsification.

\bibliographystyle{unsrt}
\bibliography{references}
\end{document}